\documentclass[sigconf,balance]{acmart}

\usepackage{tikz}
\usepackage{amsmath}
\usepackage{todonotes}
\usepackage[framemethod=TikZ]{mdframed}

\usepackage{hyperref}
\usepackage{cleveref}

\mdfdefinestyle{MyFrame}{%
  backgroundcolor=black!10,
  skipabove = 5pt,
  roundcorner=3pt
}

\newenvironment{myitemize}
  {\begin{list}{$\bullet$}{
     \setlength{\leftmargin}{1em}
     \setlength{\labelsep}{0.5em}
     \setlength{\labelwidth}{0.7em}
     \setlength{\topsep}{0.2em}
     \setlength{\partopsep}{0.2em}
   }}
  {\end{list}}

\newif\ifdraft

\drafttrue
\ifdraft
\newcommand{\BS}[1]{\todo[inline, color=orange]{BS: #1}} 
\newcommand{\FH}[1]{\todo[inline, color=blue!40]{FH: #1}} 
\newcommand{\RM}[1]{\todo[inline, color=purple!40]{RM: #1}} 
\newcommand{\TD}[1]{\todo[inline, color=green!40]{RM: #1}} 
\newcommand{\empirical}[1]{\textcolor{blue}{#1}}
\newcommand{\hideLater}[1]{{}}

\newcommand{\revise}[1]{\textcolor{orange}{#1}}
\renewcommand{\revise}[1]{#1}

\else
\newcommand{\BS}[1]{}
\newcommand{\FH}[1]{}
\newcommand{\RM}[1]{}
\newcommand{\TD}[1]{}
\newcommand{\empirical}[1]{#1}
\newcommand{\hideLater}[1]{{}}

\newcommand{\revise}[1]{\textcolor{orange}{#1}}
\renewcommand{\revise}[1]{#1}

\fi

\newcommand{\hide}[1]{}

\newcommand{\takeaway}[1]{
	\begin{mdframed}[style=MyFrame,nobreak=true]
	\textbf{Key Takeaways:} #1
	\end{mdframed}
}

\newcommand{\takeawayWO}[1]{
	\begin{mdframed}[style=MyFrame,nobreak=true]
	#1
	\end{mdframed}
}

\newcommand{\itembox}[1]{
	\begin{mdframed}[style=MyFrame,nobreak=true]
	#1
	\end{mdframed}
}

\usepackage{epigraph}

\usepackage{etoolbox}
\makeatletter

\newlength\epitextskip
\pretocmd{\@epitext}{\em}{}{}
\apptocmd{\@epitext}{\em}{}{}
\patchcmd{\epigraph}{\@epitext{#1}\\[0.5em]}{\@epitext{#1}\\[\epitextskip]}{}{}
\makeatother

\usepackage{tabularx}
\usepackage{tabulary}
\usepackage{paralist} 

\usepackage{booktabs} 
\usepackage{color, colortbl}
\usepackage{multirow}
\usepackage{array}
\usepackage{makecell}  
\usepackage{balance}
\usepackage{longtable}
\usepackage{subcaption}
\usepackage{multicol} 

\newcommand*{\tabindent}{\hspace{3mm}}

\usepackage{csquotes}

\newcommand{\quo}[1]{\textcolor{black!90}{\enquote{\emph{#1}}}}

\expandafter\def\expandafter\UrlBreaks\expandafter{\UrlBreaks
  \do\a\do\b\do\c\do\d\do\e\do\f\do\g\do\h\do\i\do\j%
  \do\k\do\l\do\m\do\n\do\o\do\p\do\q\do\r\do\s\do\t%
  \do\u\do\v\do\w\do\x\do\y\do\z\do\A\do\B\do\C\do\D%
  \do\E\do\F\do\G\do\H\do\I\do\J\do\K\do\L\do\M\do\N%
  \do\O\do\P\do\Q\do\R\do\S\do\T\do\U\do\V\do\W\do\X%
  \do\Y\do\Z}

\copyrightyear{2026}
\acmYear{2026}
\setcopyright{cc}
\setcctype{by}
\acmConference[CCS '26]{Proceedings of the 2026 ACM SIGSAC Conference on Computer and Communications Security}{November 15--19, 2026}{The Hague, Netherlands}
\acmBooktitle{Proceedings of the 2026 ACM SIGSAC Conference on Computer and Communications Security (CCS '26), November 15--19, 2026, The Hague, Netherlands}
\acmDOI{10.1145/3830454.3832690}
\acmISBN{979-8-4007-2871-6/2026/11}

\begin{document}

\date{}

\title{\underline{R}eflection, \underline{E}ducation, \underline{C}onsistency: Towards Best Ethics Practices At Security And Privacy Conferences}


\author{Florian Hantke}
\email{floria.hantke@cispa.de}
\orcid{0000-0002-5403-3524}
\affiliation{%
  \institution{CISPA Helmholtz Center for Information Security}
  \country{}
}

\author{Rafael Mrowczynski}
\email{mrowczynski@cispa.de}
\orcid{0009-0007-3877-5433}
\affiliation{%
  \institution{CISPA Helmholtz Center for Information Security}
  \country{}
}

\author{Til Dralle}
\email{til.dralle@cispa.de}
\orcid{0009-0009-7828-9899}
\affiliation{%
  \institution{CISPA Helmholtz Center for Information Security}
  \country{}
}

\author{Ben Stock}
\email{stock@cispa.de}
\orcid{0000-0001-9659-0700}
\affiliation{%
  \institution{CISPA Helmholtz Center for Information Security}
  \country{}
}

\begin{abstract}

Research ethics is a controversial and emotionally charged topic in the security and privacy community, sparking discussions at conferences and on social media. In recent years, some of the \revise{leading} conferences have introduced interventions such as mandatory ethics sections, with mixed reactions within the community. Program committee chairs and steering committees increasingly emphasize ethics, yet \revise{there is limited} empirical validation on ethics procedures and interventions, as well as \revise{no} explicitly communicated goals.

To support a shared understanding in our community and guide informed decisions at the conference level, we examined past ethics policies at the top-four conferences and conducted in-depth semi-structured interviews with senior and junior (n=20) community members, including some (former) chairs of program and research ethics committees. In these, we explored reasons for and goals of ethics procedures and discussed existing ethics interventions as well as possible new approaches proposed by our team.

While the community \revise{appears to be evolving toward the identified goals}, i.e., raising ethical awareness and preventing the publication of unethical work, we identified the lack of ethical education within the community itself as a major obstacle to further progress on this matter. Ethical education is often passed to the universities. Interviewees also mentioned challenges with the complexity of ethics policies and the lack of consistency across different conferences, as well as policies changing year-to-year. Several participants warned that over-regulation may lead to backlash, encouraging mere compliance, such as people turning to LLMs for ethics sections rather than undertaking in-depth consideration. In response, we \revise{suggest} a coordinated, community-wide ethics steering effort and take a first step by introducing an open ethics wiki.

\end{abstract}


  
\maketitle

\section{Introduction}

Interest in ethical practices has been increasing in recent years. \revise{This development follows in response to controversial incidents such as the Encore paper~\cite{burnettEncoreLightweightMeasurement2015}, which involved censorship measurements that raised concerns about risks to users in countries with repressive governments. Another example is the University of Minnesota case (also known as Hypocrite Commits)~\cite{hypocriteCommits:statement}, in which researchers intentionally submitted vulnerable code changes to the open-source Linux kernel without prior consent, resulting in objections from the affected community}. These cases have raised ethics concerns with security and privacy (S\&P) research beyond human subject studies, extending into areas such as vulnerability research, \revise{including} coordinated disclosure, and others with potential harms caused by research activities or \revise{published} outcomes. In response, conferences introduced ethics \revise{policies}, Research Ethics Committees (REC), and increasingly often \revise{require} ethics \revise{statements}. Despite these \revise{development}, the top-four conferences -- USENIX Security, IEEE S\&P, ACM CCS, and NDSS -- are still \revise{in the process of} refining balanced and consistent ethics standards. 
Currently, their approaches remain unaligned, \revise{and} differences in requirements can create inconsistent expectations across venues.
This culminated in mandatory ethics sections for USENIX Security 2025~\cite{USENIXSecurity252024}, which caught many authors off guard and led to an over 20\% rejection rate in its first cycle, primarily due to missing ethics sections~\cite{Usenix2025Message}.

A key challenge to consistent standards is the absence of well-defined and explicitly communicated goals and shared understandings of what ethical standards should achieve.
Defining a set of goals is essential to evaluate existing procedures and \revise{support the further development of institutionalization of research ethics}. Such alignment also supports more consistent and transparent decision-making across conferences, reducing the uncertainty for researchers.

\revise{We addressed this gap in the research by explicating goals and evaluations of the implemented ethics interventions. To that end, we first conducted qualitative content analyses of both the call for papers and the review forms of the top-four conferences, and combined the findings with 20 semi-structured interviews with senior and junior members of the S\&P community. In addition, we run a keyword analysis of academic publications from these conferences, covering all available proceedings from 2000 up to and including 2024.} This allowed us to explore how ethics practices evolved, what motivated their implementation, and how they help to achieve the identified goals.
%
Specifically, we answer the following: 

\begin{myitemize}
\item \revise{RQ1: \emph{How did \revise{ethics-fostering} practices of the S\&P community evolve over time -- especially at their conferences ?}}

\item RQ2: \emph{What do junior and senior members of the S\&P community perceive as the primary goals of ethics-fostering practices at major (top-four) S\&P conferences?}


\item RQ3: \emph{What ethics-fostering interventions (a) are currently in place, and (b) which additional interventions are desired by members of the S\&P community to meet identified goals?}


\item RQ4: \emph{How can conferences improve their ethics-fostering interventions in accordance with the goals and needs as perceived by community members?}
\end{myitemize}

\revise{As a result, we identified four goal sets and found} that while interventions address external impact and perception, community awareness remains an area for improvement. While mandatory ethics sections (with minor adjustments) \revise{were discussed as a step toward addressing this awareness gap, we also suggest exploring shared, community-wide ethics policies and educational materials coordinated through an ethics steering body}. As a first step, we thus launch an ethics wiki~\cite{ethicswiki} \revise{to provide information} about the topic and discuss new ethics-fostering policies. 
In summary, we make the following key contributions:



\begin{myitemize}
\item We present a qualitative analysis of ethical-reviewing practices and their development over time based on \revise{a keyword analysis in published S\&P literature,} a content analysis of 99 CfPs, and 20 semi-structured interviews with members of the S\&P community.
\item We provide a collection of ethics-fostering interventions together with their strengths and limitations based on the qualitative data \revise{indicating} interviewees' justifications.
\item We propose best-practice recommendations for improving ethics-fostering interventions and outline a path forward for community-driven development of ethical practices through an open wiki~\cite{ethicswiki}.
\end{myitemize}

\section{Background \& Related Work}
\label{sec:background}

This section offers a brief overview of \revise{the} terminology we use in this paper, \revise{background about} research ethics, and related work. It serves as a foundation for the upcoming discussion. 

\subsection{Terminology}
\label{sec:background:taxonomy}
In this paper, we discuss general \emph{ethics-fostering practices} at the leading S\&P conferences. As such, we define all activities of individual or collective actors within conference contexts in which they consider, discuss, and evolve the ethical aspects of S\&P research. 

These practices are implemented in the form of \emph{research ethics-fostering interventions} which constitute specific institutionalized and formalized efforts to modify patterns of actions, specifically behavior or decision-making regarding ethics in \revise{the S\&P} research community. They include formal mechanisms and policies introduced by conferences to promote ethical reflection and accountability, e.g., a section in calls for papers, research ethics committees, or standardized review forms that include ethics-related items.

For review forms, we speak of \emph{ethics-fostering instruments} as concrete operational tools, such as evaluation scales, qualitative feedback fields, and ethics information for authors and reviewers.

\revise{For simplicity, we use the terms \emph{ethics interventions}, \emph{ethics instruments}, and \emph{ethics} to refer to research ethics in this paper.}

\subsection{Ethics Research in Security and Privacy}

We analyze the development and operationalization of ethics interventions within S\&P conferences \revise{and therefore focus on the four highest-ranked venues according to the CORE ranking~\cite{coreranking}: USENIX Security, IEEE S\&P, ACM CCS, and NDSS}.
Interest in and the amount of more specific considerations regarding ethics has been growing in the community over the past years, as demonstrated by a growing number of ethics references in papers (\Cref{sec:histroy}) as well as by work published on this topic in recent years.

The first papers discussing research-ethics concerns and proposing corresponding standards came from the network community \revise{in 2009, following controversy within this sub-community about how to handle botnet research, e.g., whether it is acceptable to infiltrate botnets}~\cite{dittrichHaveWeCrossed2009, dittrichCommunityStandardsEthical2009}.
This early work and some informal debates surrounding it led to the Menlo Report~\cite{baileyMenloReport2012} and its companion~\cite{dittrichApplyingEthicalPrinciples2013} \revise{published in 2012 and 2013}. It is a major framework document promulgating ethical principles for all ICT studies, including S\&P research, and is still referenced today.

Since the early papers in the network research community, the interest in ethics has further increased within the S\&P community. The body of domain-specific work is growing in areas such as network measurements~\cite{partridge2016ethical, pauleyUnderstandingEthicalFrameworks2023}, server-side scanning research~\cite{hantkeWhereAreRed2024}, and censorship research~\cite{jonesEthicalConcernsCensorship2015}. The community has also discussed ethical implications of using datasets of illicit origin, such as password leaks~\cite{thomas2017ethical}. Ethical considerations emerged even in areas where they may not be self-evident, \revise{e.g., concerning} cryptographic research~\cite{rogawayMoralCharacterCryptographic2016}.
These papers often argue for controlled pre-tests before experimenting on live systems, informing participants in advance and obtaining their consent, practicing mindful and minimized data collection, considering broader societal implications, and engaging in responsible, coordinated vulnerability disclosure. 

Another body of work reasons about research ethics through the lens of concrete cases. ~\citeauthor{kohnoEthicalFrameworksComputer2023}~\cite{kohnoEthicalFrameworksComputer2023} \revise{(2023)} use the philosophical construct known as \emph{trolley problem} to discuss ethical dilemmas that can arise in S\&P research, noting that there is often no consensus on morally acceptable decisions in S\&P research. \citeauthor{macnishEthicsCybersecurityResearch2020}~\cite{macnishEthicsCybersecurityResearch2020} \revise{(2020)} highlight problems affecting institutional aspects of research ethics, in particular the lack of technical expertise in IRBs. They ground their analysis in two concrete cases: the \revise{earlier mentioned} academic Encore study \revise{published in 2015}~\cite{burnettEncoreLightweightMeasurement2015} and an industry-related case \revise{around MedSec from 2016}~\cite{medsec2016}. Both research groups advocate for an active discourse regarding ethics in cybersecurity research to raise awareness of these problems.

\revise{In 2025} \citeauthor{ramuluEthicsSecurityResearch2025}~\cite{ramuluEthicsSecurityResearch2025} studied this discourse by inquiring how ethical reasoning is practiced within the S\&P research community. They analyzed how ethical considerations are reported in papers published in 2024; \revise{in a separate component of the study, they interviewed researchers about research ethics and its challenges.} Their findings indicate that ethics in S\&P research is often difficult to define, very context-dependent, and commonly guided by the ``researchers' instincts'', prior research, and personal experience. Interviewees further mentioned a lack of formal guidance about what is considered ethical and reported difficulties in identifying relevant stakeholders.
Our study builds on this work. It re-examines and contextualizes several of the findings by empirically identifying a set of perceived ethical goals for the S\&P research community. Based on this goal-centered perspective, we further discuss in our interviews new interventions aimed at structuring and advancing ethical \revise{awareness} and practice within the community.

\citeauthor{zhangEthicsSecurityResearch2022}~\cite{zhangEthicsSecurityResearch2022} \revise{(2022) analyzed} how ethical considerations are reflected in security papers, specifically focusing on discussions of IRB approvals and other references to ethics. Its authors further surveyed Chinese researchers on their understanding of ethical principles, IRBs or similar boards, and recommendations for ethical considerations in security research. They recommend domain-specific best practices, ethics courses at universities, and obtaining ethical advice from peer reviewers. However, the paper leaves some open questions, such as how to establish ethics standards within our community and how the community could provide more technical assistance to researchers in terms of ethics. We take up these questions in our interviews and address them in our study.

Several papers suggest concrete improvements to ethical practices in S\&P research. 
For example, a recurring concern in the literature is that ethical evaluation often occurs only after the research has already been conducted. Recent work (e.g.,~\citet{hantkeWhereAreRed2024} \revise{in 2024}) suggested the pre-registration of studies in analogy to an established practice in medical research or social sciences. This idea is further developed and discussed by \citet{dirksenDontPatchResearcher2024} \revise{(2024)}, who propose Federated Review Boards as a practical mechanism for implementing early-stage ethical review.
Another concrete improvement, suggested by \citet{reidsma2023operationalizing} \revise{in 2023}, is a common coordinated vulnerability disclosure guideline in the form of self-assessment questions that could be directly adopted by Ethics Review Boards.

While prior work has discussed challenges of and proposed improvements for ethical research practices, we still miss an explication of the goals these practices are intended to achieve within the S\&P research community, as well as an evaluation of ethics-related procedures \revise{designed and implemented by }conference organizers.

\section{Methods}
\label{sec:methods}
Answering our research questions required a combined analysis of \revise{three} qualitative datasets: \revise{(1) a \emph{keyword analysis} of all top-four conference papers reaching back to the early ethics considerations to answer RQ1; (2) a \emph{qualitative-content analysis of process-data}, i.e., all publicly available calls for papers and review forms from top-four S\&P conferences, necessary to address} RQ3(a); and (3) \emph{semi-structured interviews} with senior and junior members of the S\&P research community necessary for addressing RQ2, RQ3(b), and RQ4. The following subsections outline our procedures for collecting and analyzing these datasets.

\subsection{Keyword Analysis}
\label{sec:methods:keywords}

\subsubsection{Data collection}
\revise{The number of papers at a conference mentioning ethics in their main text body can tell a lot about the general relevance of ethics at that time. We therefore leveraged DBLP and collected papers from the top-four S\&P conferences, namely ACM CCS, IEEE S\&P, NDSS, and USENIX Security, from the period 2000 to 2024. We decided to begin in 2000 to allow some time before the first ethics discussion we identified in S\&P literature (i.e., the botnet debates in the network community), while still ensuring reliable data, as papers before 2000 are often no longer accessible.
We omitted 2025 because USENIX Security began requiring ethics discussions in all papers that year, whereas before writing about ethics depended solely on the authors' motivation to do so and, possibly, on requests from specific reviewers.
After collecting the papers, we used Docling~\cite{Docling} to parse the documents into raw text.
}

\subsubsection{Data analysis}
\revise{For the keyword analysis, we created three keyword lists on different topics. \emph{General Ethics} including general references to ethical aspects (e.g., review board, ethical concerns; see \Cref{tab:keywords}); \emph{Human Subjects} covering references to studies with human participants (e.g., users or developers); and \emph{Botnet} containing keywords typical in botnet research\footnote{\revise{We initially planned to include the trends around human subjects and botnet research to analyze potential correlations within topics. However, due to a range of limitations, we decided to use only the analysis of the ethics mentions as a motivational input and mention the additional graphs in \Cref{appx:keyword}.}}.}

\revise{
To create these keyword lists, we collected text from chapters filtered by their chapter headings (see~\Cref{tab:keywords}). For instance, for the \emph{General Ethics} topic, we focused only on chapters whose headings contained the word \emph{ethic}, including its various variations.
We used these texts as a base to extract the final keyword lists (see~\Cref{tab:keywords}) by leveraging statistical features such as TF-IDF weighting (i.e., term relevance) and word co-occurrences, followed by manual curation.
These widely used techniques in keyword extraction~\cite{firoozeh2020keyword} help ensure \emph{representative} keyword sets; they enable a more robust keyword analysis (i.e., fewer false positives) compared to prior work (e.g.,~\cite{zhangEthicsSecurityResearch2022}), as it relies on groups of related terms rather than individual keywords.
To measure the presence of the topics over time, all papers containing at least one term from the corresponding keyword list are counted in.
}


\subsubsection{Limitations}
\revise{
While a keyword analysis seems suitable for assessing the relevance of ethics within the S\&P community over time, it also has a number of limitations. We therefore use this analysis only as an additional motivational part rather than the foundation of our argumentation. Nevertheless, we still consider the findings to be worth reporting.
}

\revise{
One of the major problems with the keyword analysis, as carried out here, is that the results depend on the quality of the keyword lists used. Although TF-IDF and word co-occurrences can ensure a certain quality in terms of representativeness and specificity of the keywords, the use of these methods does not provide clear results. For example, one particularly representative word pair for texts whose headline contains \emph{botnet} is \emph{IP addresses}. However, this word pair can be found across many areas of computer science. The approach used here cannot compare different thematic areas of computer science in order to further increase the specificity of the keywords and identify clearly distinct terms.
}

\revise{
Another limitation is that the keyword-based approach can be influenced by confounding effects. An increase in the number of papers mentioning ethics over time does not necessarily reflect a shift in the community, but may instead be driven by a growing share of papers in more ethics-aware subfields, such as human-centered security. 
Second, the way the community might react to the emerging treatment of ethical issues can affect the trend. It is possible, for example, that there is a delay in responding to novel interventions such as the Menlo Report or that conferences will influence each other in their approach to the topic of ethics, which, however, only later reflects in papers. This makes it extremely difficult to trace the results back to specific historical events.
}

\revise{
Furthermore, the method used here is not suitable for measuring arbitrary topics. We made attempts to create keyword lists for the areas of \emph{attack} and \emph{measurement} research. However, the chapters titled with the respective terms are far too numerous and cover too many subtopics to extract a specific keyword list.
To use this method, a more precise understanding of potential sub-communities and sub-topics must be available in advance.
}

\subsection{Process-Data Analysis}
\label{sec:methods:content}

\subsubsection{Dataset and Analytical Procedure}
A Call for Papers (CfP) states the conference's topical goals and defines the content-related and format expectations of conference organizers. 
CfPs can thus be considered historic documents \revise{that} define key parameters for conference publications. \revise{In other words, they shape processes of paper selection at their initial stage. Social scientists call it process data, because this data is generated by the very processes that are studied~\cite{baur2011mixing, grenz2020processualizing, schmitz2022data}.}
Against this background, our general RQ3(a) can be specified in the following way: How had the expectation of adherence to ethical norms been communicated in CfPs of the top-four conferences, and what specific ethics interventions have been stipulated in these documents?

Against this backdrop, we analyzed all publicly accessible CfPs and submission guidelines of the top-four S\&P conferences: CCS, IEEE S\&P, NDSS, and USENIX Security (see \Cref{tab:cfps}). \revise{USENIX Security, in addition to the CfPs, also publishes an annual message from the PC Co-Chairs. We also included these messages in our dataset, as they provide context and justification for the chair's decisions.} Two researcher analyzed all documents (CfPs $n=99$; Chair messages $n=12$) employing qualitative content analysis~\cite{Schreier.2014, Mayring.2000} by coding each occurrence of ethics-related content. \revise{The initial codebook was developed by the lead author in a bottom-up manner and refined in discussion with the second coder (codebook in~\Cref{appx:process-data-codebook}). Their independent codings reached a Holsti Index~\cite{mao2017intercoder} (i.e., intercoder agreement) of 96.9\%. After this comparison, both coders clarified disagreements until they reached consensus}. As a final result, they identified various forms of ethics interventions.

\begin{table}[]
    \centering
    \footnotesize	
    \begin{tabular}{l l l l l l l}
\bf Conference & \bf Number & \bf \revise{Conference Years}\\ \toprule
USENIX CfP & 28 & 1998 - 2025 \\
USENIX Chair's Message & 12 & 2013 - 2025 \\
USENIX Review Form & 16 & 2019 - 2024 \\ \midrule
NDSS CfP & \revise{25} & 2001 - 2025 \\
NDSS Review Form & \revise{10} & 2020 - 2025 (missing '23) \\ \midrule
S\&P CfP & 24 & 2002 - 2025 \\ 
S\&P Review Form & 2 & 2025 \\ \midrule
CCS CfP & 22 & 2004 - 2025 \\
CCS Review Form & 9 & 2019 - 2024 (missing '23) \\
\bottomrule
\end{tabular}

    \caption{Document corpus overview. Review forms include multiple cycles per year.}
    \label{tab:cfps}
\end{table}

In addition to publicly available materials, we analyzed review forms \revise{($n=37$)} dating back to 2019. We collected these documents by asking senior community members to share documents from conferences for which they served on the PC \revise{and providing them a script that enabled the anonymized extraction from HotCRP. As the S\&P HotCRP instance is only available from 2025 onward, the dataset for this venue contains only forms from this year.} The lead researcher then used qualitative content analysis to compile a list of ethics instruments \revise{using the codebook in~\Cref{appx:review-codebook}.} 


\subsubsection{Limitations}

Our analysis primarily focuses on the S\&P community. Although we surveyed a range of related fields, other scientific communities may have developed effective or innovative interventions that we were not able to identify. Thus, our proposed interventions may be only a subset of possible options. Nevertheless, we consider this selection a valuable starting point for a broader discussion about additional ethics interventions we want to deploy.


\subsection{Interviews}
\label{sec:methods:interviews}

Process data \revise{from the CfP dataset} is an excellent source of information about official stipulations and procedures (e.g., the introduction of RECs). \revise{However, they do not capture} informal, often orally performed, activities or motivations behind certain publicly visible decisions. Since we are also interested in goals of ethics interventions, as perceived by S\&P community members, and their assessments of existing and possible new interventions, we conducted semi-structured interviews~\cite{Flick.2023, Roulston.2018}, drawing on preliminary findings from our process-data analysis. These interviews can be characterized as expert interviews~\cite{Bogner.2009}, because we aimed to \revise{draw on} participants' first-hand experiences with various aspects of ethics-fostering practices \revise{including} changes over time.

\subsubsection{Interview Guide}

We developed \revise{our} interview guide, drawing on the list of ethics interventions identified in the process-data analysis. It consisted of seven thematic blocks, each containing general and, where appropriate, follow-up questions addressing the following topics: (1) \emph{ethics practices and reviewing procedures}, (2) \emph{existing ethics interventions}, (3) \emph{proposed ethics interventions}, (4) \emph{ethics instruments}, the (5) \emph{reasoning} and the (6) \emph{goal} behind ethics practices, and the (7) \emph{importance} of ethics. For the proposed ethics interventions and instruments, we used a survey instrument (see \Cref{sec:methods:survey}) embedded in our interview guide. We also invited participants to share any additional thoughts. The complete base interview guide is provided in~\Cref{appx:interview-guideline}. We adapted it slightly for different categories of research participants.

We conducted five pilot interviews to test our interview guide \revise{with participants of differing backgrounds, given the limited pool of real participants}: three with former PC chairs and two with doctoral researchers. \revise{As these pilot interviews did not reveal any major issues, we continued the proper interviews using the same guide. After the first two regular interviews, we}, however, added an explicit question block about existing ethics interventions; earlier, this was covered only implicitly via questions about ethics procedures. 

\subsubsection{Survey Instrument}
\label{sec:methods:survey}
Research ethics and corresponding interventions are very abstract topics. Thus, we expected less topic-aware participants to struggle articulating their views and reasoning. Hence, we incorporated a practical survey instrument into the interview to encourage more concrete statements.

Based on the process-data analysis, we identified potential improvements of ethics interventions beyond those already in place: (1) for \emph{research ethics committees} to \emph{consolidate RECs of different conferences into one inter-conference body}, \emph{include external experts}, and implement a \emph{REC-member mentoring system}; (2) \emph{feedback to the authors} through \emph{ethics meta reviews} and a \emph{``profound ethical considerations'' badge};  and (3) \emph{education} with \emph{educational material in form of categories} and \emph{ethical training for reviewers}. This list of new interventions is not meant to be complete, but instead to gather first insights into interviewees' reactions to such innovations.

The survey instrument consisted of two pages (see an example and all items in \Cref{appx:interview-guideline}). The first page presented the proposed interventions sequentially and in random order, and asked to rate each item using a five-point Likert scale (strongly agree to strongly disagree).
The second page listed items for review forms: \emph{ethics scores}, \emph{feedback to authors} fields, and \emph{internal discussion} options. Those were also rated on a five-point Likert scale. 

\subsubsection{Sampling}
We specifically targeted members of the \revise{S\&P} community \revise{and}
 deliberately selected participants differing in their professional seniority: (1) very experienced community members who have served as PC or REC chairs and thus played central roles in shaping conference policies; (2) junior researchers, defined as paper authors without prior chairing experience. To ensure a diversity of perspectives, we further considered research background, gender, and nationality in our sampling procedure.

\subsubsection{Recruitment}
We used two different recruitment strategies. As for the senior members, we invited, on the one hand, individuals identified in our process-data analysis as directly involved in implementing significant procedural or structural changes \revise{at one of} the top-four S\&P conferences. On the other hand, to cover a broad spectrum of opinions, we also included individuals who kept ethics procedures unchanged when they were chairing a conference. \revise{All e-mails were sent by the lead author, a doctoral researcher (see~\Cref{appx:recruitment:senior} for the message).}
For the junior members, we examined all papers published at the top-four conferences in 2025 and assigned them to research fields (see~\Cref{appx:participants} for details). We then randomly selected one paper from each field and invited the first author of each paper. In cases where the first author was not a doctoral candidate, we invited the first doctoral candidate listed among the authors. This approach allowed us to recruit a heterogeneous group of early-career researchers across various research areas. To minimize unsolicited communication, \revise{our lead author} invited these \revise{potential interviewees} by e-mail in batches of ten \revise{(see~\Cref{appx:recruitment:junior} for the e-mail)}. Per \revise{person}, only one reminder \revise{was sent} after roughly a week if they did not respond. We subsequently proceeded with the next batch until we reached the coverage of nearly all research fields.

\begin{table}
    \centering
    \footnotesize	
    \begin{tabular}{l l l l l}
     ID & Seniority & Service Roles & Length & Training \\ \midrule
     I-01 & Senior & PC-Chair & 01:12 & \\ 
     I-02 & Senior & PC-Chair, REC-Chair & 00:56 & \\ 
     I-03 & Senior & REC-Chair & 01:28 & Yes \\ 
     I-04 & Junior & PC-Member & 01:05 & \\ 
     I-05 & Junior & - & 01:26 & Yes \\ 
     I-06 & Junior & - & 01:29 & Yes \\ 
     I-07 & Junior & - & 00:55 & \\ 
     I-08 & Senior & PC-Chair & 01:36 \textcircled{2} & Yes \\ 
     I-09 & Senior & PC-Chair & 01:06 & \\ 
     I-10 & Junior & - & 00:42 & \\ 
     I-11 & Junior & - & 01:00 & \\ 
     I-12 & Senior & PC-Chair & 01:09 & \\ 
     I-13 & Senior & Track-Chair & 01:50 \textcircled{2} & Yes \\ 
     I-14 & Junior & - & 01:08 & \\ 
     I-15 & Junior & - & 01:09 & \\ 
     I-16 & Senior & PC-Chair, REC-Chair & 01:10 & Yes \\ 
     I-17 & Junior & - & 01:01 & \\ 
     I-18 & Senior & PC-Chair & 01:03 & Yes \\ 
     I-19 & Senior & PC-Chair & 01:27 \textcircled{2} & \\ 
     I-20 & Junior & - & 00:52 & \\ 
     \bottomrule
     & Average & - & 01:11 & \\
\end{tabular}
    \caption{Interviewees' backgrounds, interview length, and whether they have received ethical training before. Interviews marked with \textcircled{2} were conducted in two parts.}
    \label{tab:interviewees-body}
\end{table}

\revise{\Cref{tab:interviewees-body} shows, resulting from the recruitment effort,} we interviewed \empirical{10} senior members of the S\&P community, among them multiple PC chairs and REC chairs, and \empirical{10} junior members who are mostly PhD candidates (with \empirical{one} having served on a PC).
As the pool of potential interviewees is very small, we decided not to disclose further details on individual participants \revise{beyond their service roles and whether they have reported having received prior ethical training (e.g., through an ethics board or university course),} in order to preserve their anonymity. Yet, we recruited interviewees from various regions, including North and South America, Europe, and Asia \revise{(as detailed in \Cref{appx:participants})}. We also included a wide range of research areas, covering all of our sampling matrix with the exception of \emph{Wireless Security} (see ~\Cref{appx:recruitment:senior}).

\subsubsection{Interviewing Process}
Before the interview, every participant filled out an \revise{online} questionnaire about their research background, \revise{ethics training (if ever taken),} and conference services; \revise{it was combined with the informed-consent form (see \Cref{appx:consent}}).

Interviews lasted between 42 and 110 minutes. \revise{The lead author (without the involvement of the PI)} interviewed between late April 2025 and early January 2026. All interviews were conducted and recorded via Zoom. After transcribing, we deleted the video file and only kept the audio for better coding flows. For transcribing, we used a local tool based on OpenAI's Whisper. A student assistant proofread and anonymized all transcripts. \revise{In the end, we also deleted all audio files and kept only the transcripts.}

\subsubsection{Data Analysis}
We conducted a qualitative content analysis~\cite{Schreier.2014, Mayring.2000} of our interview data \revise{(not including the pilot interviews)} in a flexible manner: on the one hand, we used a pre-defined set of deductive codes derived from the literature on ethics in S\&P research and our interview guide, but on the other hand, we were open to new data-driven insights captured as inductive codes created in a bottom-up manner (codebook in~\Cref{appx:interview-codebook}).

Our research team consisted of two \revise{cybersecurity} researchers and two social scientists. At the initial stage, the lead author (cybersecurity researcher) and one of the social scientists \revise{(second author)} coded five interviews (using ATLAS.ti CAQDAS package) and discussed the resulting codings within the team. In this process, we realized that both coders had coded very similarly with pre-existing deductive codes. They achieved high inter-coder agreement, with a Holsti Index~\cite{mao2017intercoder} of 93.3\%. As for the newly proposed inductive codes, both researchers discussed them until they resolved all divergences and reached intersubjective agreement on their use~\cite{mcdonald2019reliability, wermke2022committed}. 

After compiling an integrated codebook, the lead author continued with coding of the entire interview dataset. \revise{Simultaneously, the other social scientist (third author) coded all interview transcripts with deductive codes. This additional coding procedure aimed at testing whether a researcher who, up to this point, was not involved in the transcript analysis would similarly apply our deductive codes. The comparison of the codings by these two team members who coded all transcripts resulted in a high inter-coder agreement (Holsti Index = 93,7\%).}

At the interpretation stage, two team members \revise{(the first author and the second author)} extracted relevant information from transcripts using the shared code system and entered this information into tables, which then allowed them to compare individual interviews along key categories (e.g., understanding of goals, assessment of specific interventions, etc.). Next, they identified and systematized commonalities as well as differences in statements made by different participants. The final result was a systematic description of the spectrum of opinions articulated by interviewees.

\subsubsection{Interview Limitations}
\label{sec:methods:limitations-interview}
Since it is a qualitative interview study, the frequency of opinions or judgments expressed by our participants must not be interpreted as indicative of frequency distribution within the entire population of S\&P researchers.
Our interview dataset can also be affected by the self-selection bias, skewing the interviewee sample towards those who are more ethics-sensitive and hence more willing to participate in such a study.
Another limitation stems from \emph{sampling bias}, as we targeted popular S\&P conferences and authors of published papers. As a consequence, we missed researchers who never published there -- some of them possibly due to ethical reasons.
\revise{Additionally, by focusing on junior and senior members, our sampling strategy excludes the opinions of mid-career community members, e.g., people serving on a PC without holding a chair role.}
We also assume that a study on a sensitive topic such as research ethics could be influenced by the \emph{social desirability bias}. Furthermore, we told our interviewees to ignore feasibility, which probably made them more open to additional interventions, because resource considerations were faded out. 

\subsection{\revise{Further} Limitations}

\revise{
Besides the methodological limitations discussed before, this study is also limited by its focus on the top-four conferences: USENIX Security, IEEE S\&P, ACM CCS, and NDSS. We decided on this scope because we view these conferences as influential in shaping community norms, often with downstream effects on other venues. At the same time, this focus may overlook ethical practices applied by smaller or more specialized conferences, where interventions may already be more advanced. Nevertheless, our goal is to capture perspectives at the level of the broader S\&P research community, for which these top-four conferences are representative.
}
\section{Ethics at the Top-four Conferences} 
\label{sec:histroy}

\begin{figure}
    \centering
    \includegraphics[width=0.9\linewidth]{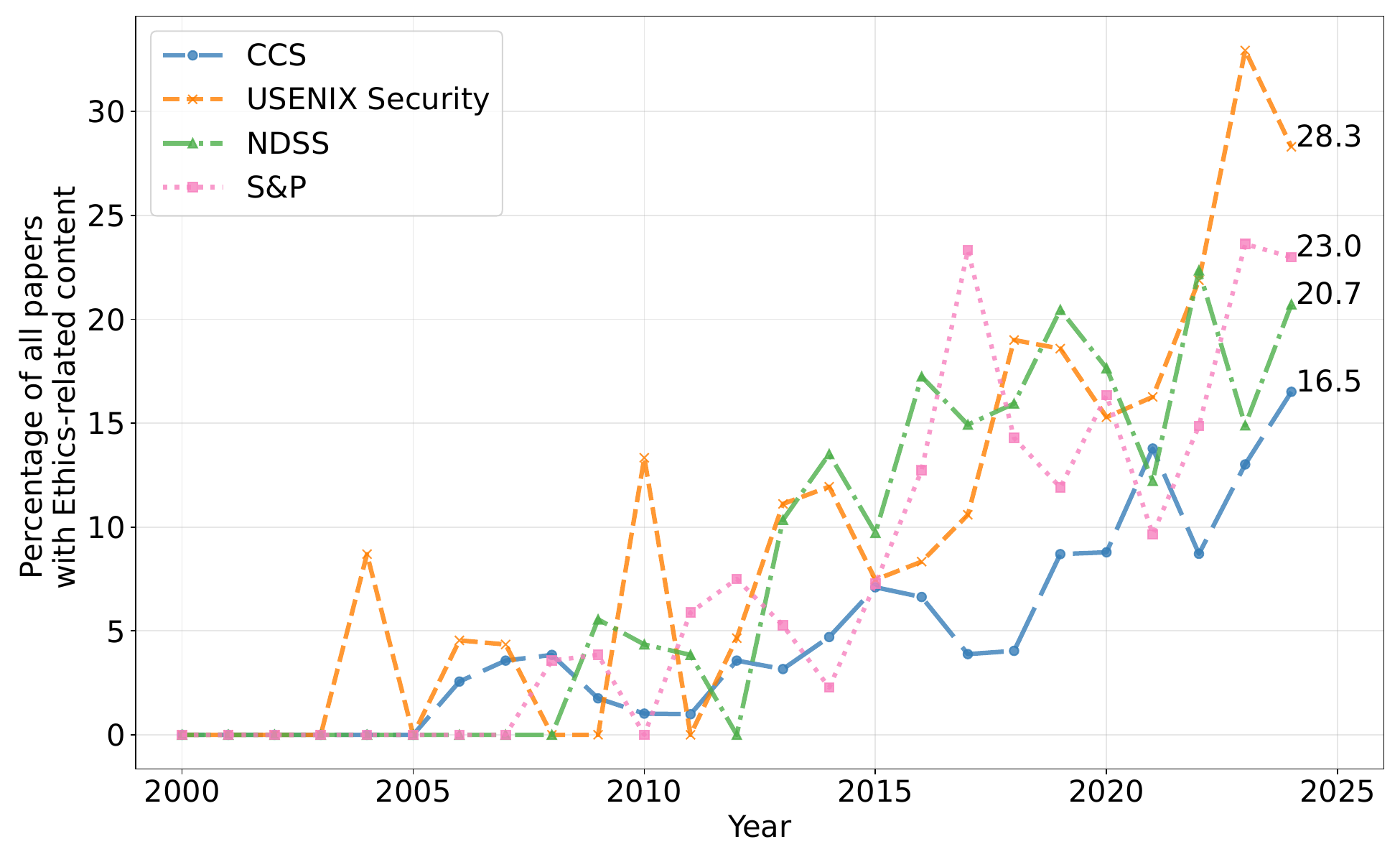}
    \caption{Temporal distribution of papers containing ethics-related content, separated by conference venue.}
    \label{fig:main:general_ethics_over_time}
\end{figure}


One of our interviewees pointed out that in S\&P research, \quo{ethics has always been important, but not everyone understood, not everyone believed it was important, or knew that it was important.}\hideLater{(19:74)} As of today, ethical considerations have found their way into our community and its major conferences. So how has this happened? 

In the following, we present a brief history of ethics debates within the S\&P research community and the more recent introduction of various ethics interventions. We combine findings from three different sources: (a) the qualitative content analysis of process data; (b) our semi-structured interviews, especially those with senior members;
(c) a quantitative keyword analysis of papers published at the top-four conferences. The CfP and the keyword analysis provide the factual skeleton of our presentation, while interviews capture informal processes and likely motivational structures behind the factual changes.

The keyword analysis demonstrates a significant increase in ethics-related vocabulary since the early 2010s (see~\Cref{fig:main:general_ethics_over_time}). 
While largely absent well into the late 2000s, by 2024
ethics-related keywords appeared in 22.3\% of all papers accepted at top-four conferences (CCS \empirical{16.5\%}, IEEE S\&P \empirical{23.0\%}, NDSS \empirical{20.7\%}, and USENIX \empirical{28.3\%}).
We thus ask what processes within the S\&P research community led to this significant change.

Our interviewees were unanimous in describing the trend in the ethics debates: it was a movement from informal discussions, initially held at low-profile venues like co-located workshops (e.g.,~\cite{allodi2009leet, dittrich2010building})\hideLater{(12:37,:24-27)}, towards an institutionalization of ethical reviewing and formal procedures\hideLater{(01:13; 02:98,:110; 03:149; 08:13; 16:30-31; 18:55; 19:65)} (stated by 7 interviewees). The outcomes were: (a) ethical requirements and guidelines in CfPs, (b) formalized ethical-reviewing procedures, either conducted by a conference-wide REC, as is currently the case at NDSS, IEEE S\&P, and USENIX Security, or by track chairs within the domain of their scientific expertise, as at CCS\hideLater{(09:12,:19)}, and (c) mandatory ethics sections introduced so far only by USENIX Security in 2025.

Interviewees recalled that the first very informal debates about ethical aspects started already in the 2000s. They were sparked by problems arising from research on botnets and honeypots and studies using leaked personal data from \quo{big data dumps}\hideLater{(16:24)}. At this time, major conferences were small enough for program committees to meet in-person in \quo{a conference room}\hideLater{(19:65)} and discuss every submission, during which ethical concerns were raised before accept-or-reject decisions were made\hideLater{(02:50; 12:37,:24-27; 19:65)}. Two eye-witness interviewees reported that ethics was back then \quo{an implicit question for the reviewers} \hideLater{(02:49)} and PC chairs were able to \quo{keep an eye on this thing}. \hideLater{(13:17)}

Dedicated workshops on research ethics organized by the USENIX Association around 2010 (e.g.,~\cite{allodi2009leet}) were described by one interviewee as \quo{the kick-off for the Menlo Report}\hideLater{(16:24)} -- the first milestone in institutionalizing ethics on ICT-related topics. Published in 2012, the Menlo Report was the first attempt to systematize, explicate, and promulgate an ethical framework for the entire ICT and S\&P research. It outlines some general ethical principles \revise{(although it does not systematically engage with philosophical debates on research ethics except for one general reference in one of its endnotes}, but interviewees perceived it as not very \quo{actionable} and too focused on specific problems of the time in which it was compiled\hideLater{(08:142)}. Our CfP content analysis indicates something remarkable in this context: program committees began mentioning the Menlo Report only in 2022 (IEEE S\&P), although the general topic of research ethics appeared in CfPs for the first time already in 2013. By 2025, NDSS, IEEE S\&P, and USENIX had all referenced the Menlo Report in their CfPs. This fact suggests that the debates leading to the publication of the Menlo Report were still confined to a smaller circle of ICT and S\&P researchers with a particular interest in ethics, and had not yet reached the broader S\&P community.

The first institutionalization step at the conference level was made by USENIX Security in 2013: a one-sentence reference to research ethics was included for the first time in the call for papers \revise{as both, our }process-data analysis \revise{and our interviews, indicate}) \hideLater{(see also: 16:30,:206; as well as: 01:31; 12:17; 14:37)}. NDSS followed in 2015. IEEE S\&P and CCS mentioned ethics in their 2017 CfPs for the first time.

In the early 2020s, the top-four conferences started to set up research ethics committees (RECs), initiated by IEEE S\&P and USENIX Security in 2022. Interviewees linked this trend partly to the public controversy after the acceptance of the \quo{Hypocrite Commits} paper~\cite{hypocriteCommits:statement} at IEEE S\&P 2021\hideLater{(08:13; 16:16)}. The first REC at USENIX Security 2022 was still semi-formal: 15 experienced PC members commanding scientific expertise from different research fields were asked by PC chairs to deal with potential ethical issues indicated by reviewers~\cite{Usenix2022Message}.\hideLater{(08:36-38,:42)} In the following years, the USENIX REC became more formalized\hideLater{(08:38)}. As of early 2026, IEEE S\&P, USENIX and NDSS have a REC, while there are, as of today, no signs that the CCS organizers plan to introduce a centralized REC for the 2026 edition and rather stick with its more decentralized ethical-reviewing procedure involving track chairs in pivotal roles\hideLater{(09:23)}.

According to REC chairs we talked to, \revise{such a body typically operates} as follows: first, reviewers flag some papers in their review form as requiring additional ethical considerations. Then one or two REC members inspect the paper and assess handling of ethical issues, ask questions to the authors via HotCRP \revise{(i.e., a paper submission system}), and try to clarify open ethical problems. RECs have no formal power to reject a paper, but provide recommendations, with final decisions made by the PC chairs. In some cases, those can decide to publish a paper with a statement on ethics attached~\cite{burnettEncoreLightweightMeasurement2015, solomosTalesFaviconsCaches2021}. \revise{From a philosophical perspective, we can see here a form of discourse ethics at work~\cite{apel1996_selected, habermas1990_moral}.}

Our interview dataset helps explain both key aspects of the trend described above: (1) institutionalization: Why research ethics have become so important for the S\&P community that ethics-fostering interventions were integrated into the reviewing process. (2) formalization: why reviewing procedures, including their ethics-fostering components, have become more formalized over time.

Regarding institutionalization of research ethics, our interviewees pointed to the growing ethical awareness within the community\hideLater{(02:91,:55-56; 09:55; 14:37; 19:74)}, the absence of IRBs/ERBs in many countries\hideLater{(12:11)} and an expanding presence of researchers from areas in which ethics traditionally play a significant role\hideLater{(01:9; 02:56)}. The reasons for the increased sensitivity for ethical issues were mostly seen in some controversial publications\hideLater{(09:167)}, in particular, the \quo{Hypocrite Commits} paper was mentioned most frequently~\hideLater{(01:9-11; 02:91; 05:186; 08:13; 10:146,:156; 15:189-204; 16:16; 17:186; 18:55)}(9 interviewees). However, one senior participant argued that this highly controversial paper had merely been \quo{a catalyst} of further institutionalization steps rather than the main reason. Instead, they pointed to the broader rise in ethical sensitivity across society, which affected the ethical awareness of the S\&P community in the years following the COVID pandemic and the BLM movement in the US\hideLater{(19:65)}.

As for the formalization of reviewing procedures, our interviewees pointed in particular to the numeric growth of the conferences in terms of submissions (more than doubling since 2020~\cite{stattot}) and reviewers \hideLater{(01:13; 02:50,:91,:110; 03:16; 07:204; 19:65)} (5 interviewees). 
This scale made the formalization of reviewing procedures, including their ethics-related aspects, \revise{seem} necessary to handle the \revise{exponential} growth of papers. The broadening of the topical scope of conferences \hideLater{(02:56,:59)} and their increasing socio-cultural diversity \hideLater{(07:204)} were also sometimes mentioned as a reason for formalization, which allows for a higher transparency of procedures necessary when community members differ in their understandings of research ethics. 

\section{Goals of Ethics Interventions}
\label{sec:goals}

The assessment of ethics interventions, present as well as potential/future, requires an estimation of goals supposed to be achieved by these interventions as part of the overall reviewing practice. Hence, we examined our interviewees' explicit statements on the goals of ethics interventions and the progress towards them.

\subsection{Perceived Goals}

We asked all interviewees what they perceive as the main goals of ethics\revise{-fostering} interventions introduced at the major S\&P conferences in recent years. Their answers focused on four major \revise{categories}: (1) normative regulation of various implications that S\&P research activities (can) have for the outside world (on individual society members, certain social groups or organizations, society at large, etc.); (2) impact on the S\&P research community itself; (3) influence on its individual members; (4) perception and evaluation of that community by various external actors. \revise{Category two and three} are closely interrelated; hence, we discuss them in combination. Nearly all of our participants named more than one specific goal scattered across several of the above-mentioned categories.

The normative regulation of research activities with respect to their impact on the outside world was predominantly addressed by our interviewees as \emph{avoiding harm}\hideLater{(03:44; 04:94,96,104,113; 12:268; 14:325; 15:188)}. In a similar vein, one \revise{junior} participant stated that the goal was to prevent \emph{malicious behavior} of researchers\hideLater{(15:73,206)}.
Yet another \revise{junior} interviewee defined the goal of ethics interventions as \emph{protecting people} and \emph{communities}\hideLater{(17:188)}. The same participant also mentioned environmental and computational resources, as well as \emph{infrastructures}, as aspects that should also be taken into consideration when assessing potential damage inflicted by S\&P research\hideLater{(17:196)}.

Our research participants understand \emph{harm} in a two-fold way: (a) a negative consequence directly connected to research activities (e.g., exposing human subjects to excessive stress during an experiment); (b) a negative consequence resulting from the publication of research results (e.g., a study disclosing an unknown vulnerability).

\revise{Some interviewees described ethical review as \emph{minimizing harm} while weighing it against potential benefits of S\&P studies\hideLater{(01:40,42; 13:118,124-126)}—or, as one participant put it, pursuing \quo{the goal of progressing in science without creating damage}\hideLater{(06:257)} -- reflecting an implicitly consequentialist, utilitarian view of research ethics~\cite{Bentham.1970, kohnoEthicalFrameworksComputer2023}. By contrast, interviewees who emphasized only \emph{avoiding harm} could be read either consequentialist or deontological~\cite{Kant.2011, kohnoEthicalFrameworksComputer2023}, since \emph{harm} may also denote a rights violation that cannot be traded off against benefits, especially benefits to others or \emph{society at large}.}

The impact of ethics interventions on the S\&P community itself was most commonly framed as \revise{raising} the ethical awareness within it\hideLater{(03:154-155; 05:188; 07:206; 10:136; 11:41,130,132; 14:325; 15:134; 16:50)}, making authors think about ethical aspects of their scientific work\hideLater{(01:59; 02:125; 03:50-52,149; 04:170; 08:51; 11:132; 13:46; 14:79,329; 16:50; 17:21; 18:55)} and \quo{educating} them\hideLater{(05:188; 09:55)}. One \revise{PC-Chair} even spoke of some ethics interventions as \quo{a pedagogical exercise in terms of helping people learn how to do research in more ethical ways}\hideLater{(18:55)}. These interconnected aspects were most frequently referred to in the context of mandatory ethics sections introduced at USENIX Security 2025.

Complementary framings focused on the impact of ethics interventions on the S\&P community, emphasizing that it helped develop common community norms\hideLater{(08:17; 09:169; 19:84)}, making them more explicit\hideLater{(02:91,98; 03:15319:84)} and increasing the coherence of ethics-related decisions\hideLater{(02:60; 03:149; 08:47,49)}. One \revise{senior} interviewee mentioned that institutionalized ethical reviewing can help to integrate a growing and diversifying S\&P research community around core values\hideLater{(19:84)}. Another \revise{senior participant} added that it produces documentations which will allow to trace how ethical understanding and awareness of the community evolves over time\hideLater{(02:125)}.

Furthermore, enforcement of ethical norms within the S\&P community by denying publication of papers presenting research deemed as unethical was mentioned as an important goal specifically linked to the \revise{intervention} \emph{right to reject}\hideLater{(08:47,49; 08:72; 16:42)}.

As already mentioned, some interviewees pointed out goals primarily focused on individual community members rather than the entire community. They emphasized that formalized ethical reviewing provides guidance for authors\hideLater{(01:27; 13:104)} and reviewers\hideLater{(03:149;14:53)} and can also contribute to overall paper quality\hideLater{(01:29; 19:89)}. We see these individual-level goals very closely intertwined with community-centered goals, because behavioral patterns of authors and reviewers shape major community-wide practices. At the same time, established community norms influence the behavior of individual members: those who are not able or willing to adhere to present ethical norms are sanctioned because they are denied publication opportunities.

The fourth \revise{category} of goals proposed by our interviewees focused on how the S\&P research community is perceived and evaluated by external actors, including private businesses, state agencies, legal systems, and the broader public. Ethics interventions were seen as ways to aim at improving the reputation of S\&P conferences\hideLater{(07:204; 14:319; 18:55; 19:65)}, research institutions\hideLater{(11:130)} and the entire academic science\hideLater{(06:257; 15:206)}. In some extreme cases, these measures could also serve the purpose of protecting conference organizers from legal liability\hideLater{(08:133,136; 12:268; 13:118; 14:319)}.

\subsection{Progress Towards Perceived Goals}

We also asked our interviewees to assess how far the S\&P research community has progressed \revise{toward} the above-mentioned goals. A clear majority of our participants stated that the implementation of ethics interventions is an \quo{ongoing process}\hideLater{(01:44; 02:49; 03:157; 04:130; 06:259; 08:19; 09:171; 10:144; 11:132; 14:331; 15:209; 16:206; 17:202; 18:61)}. Some added that this process can never end, because it is \quo{an iterative improvement}\hideLater{(10:144)}.
Some added that there will always be something ethically incorrect happening\hideLater{(02:115)} and a need for the community to learn will persist\hideLater{(16:206)}. Nevertheless, the same research participants also \revise{noted} that recent ethics-fostering activities are a \quo{\revise{step in the right direction}}\hideLater{(07:208; 10:144; 15:209; 16:206; 17:202)}. 

One \revise{junior} interviewee presented a more ambivalent progress assessment: \revise{while ethics interventions have already been fairly institutionalized, with \quo{mechanisms to check the ethics problems} in place, there are still papers published although they \quo{have some ethics problems}}\hideLater{(07:208)}. Two other participants went even further and stated that the goals of establishing ethics-fostering procedures at major S\&P conferences have already been reached. A senior researcher \revise{pointed out}: there are IRBs, ethical guidelines, and RECs at conferences; hence, no additional institutions are needed\hideLater{(12:124)}. \revise{They noted that} many community members already consider ethical aspects of their scientific work\hideLater{(12:274)} \revise{and that} ethically questionable studies that were possible \quo{15 years ago, we can’t do them today}. In conclusion, this interviewee stated about research ethics: \quo{it’s not as big as a concern}\hideLater{(12:274)}. A junior researcher pointed to the mandatory ethics section \revise{at USENIX Security '25 to support the view that ethics goals have already been achieved}.

\revise{As this study uses qualitative methods, it does not allow us to \revise{infer} the distribution of opinions across the broader S\&P community.}
\revise{Yet,} the views presented by two interviewees before offer valuable insights into perspectives skeptical of further institutionalization beyond the status quo. Notably, the goal perceptions of both participants were focused on external impact, specifically harm avoidance\hideLater{(12:268; 20:131)}, and on protection from legal liability\hideLater{(12:13)}. They were the only two who did not mention community-centered goals, the otherwise most widely discussed set of goals among our other 18 participants.

\takeaway{We identified four goal sets \revise{(GS)}:
(1) Impact on the outside world society; (2) impact on the S\&P community; (3) influence on individual members; (4) external perception.
}

\section{Ethics-fostering Interventions}
\label{sec:measures}


In the previous sections, we described the evolution of ethics interventions at S\&P conferences and their goals as perceived by our interviewees. With these findings in mind, we now turn to interviewees' attitudes towards existing interventions, new proposals, and suggested improvements that came up during the interviews.

\subsection{Existing Interventions}
\label{sec:interventions:existing}

Over the past years, conferences have experimented with a range of ethics interventions. To get a more empirically grounded understanding of their effect on the community, we discuss the most widely adopted interventions below.



\subsubsection{Right to Reject}
\label{sec:interventions:reject}

Conferences reserve the \quo{right to reject a submission if insufficient evidence was presented that ethical and legal concerns were appropriately addressed}. 
It is the ultima ratio when no other intervention is successful.

Interviewees commonly agreed that conferences' right to reject is essential for sanctioning unethical behavior and enforcing community standards, echoing prior findings~\cite{ramuluEthicsSecurityResearch2025}.
Without rejecting papers, some authors might not care about our moral standards, because, \revise{as a former PC-Chair put it,} \quo{if no one is ever rejected for unethical behavior, then there is no incentive for people who might want to act unethically}. 
\revise{Another chair} also emphasized that rejection can serve as a learning experience, helping researchers to reflect on ethical shortcomings of their research design: 
\quo{You know people got rejected [...] it's a learning experience. It doesn't mean you're a bad person. It just means you didn't think about certain things. And then the community is telling you, you shouldn't have done this.} 

However, participants also emphasized that rejection should generally be a last resort, \revise{as a junior member said,} \quo{after a discussion with the authors and giving them an opportunity to comment.} 
Ethical concerns were described as highly case-specific, depending on the severity and irreversibility of the harm: while issues like missing coordinated vulnerability disclosure can often be resolved in rebuttal, harms such as poor participant treatment cannot be fixed post hoc. One \revise{REC-Chair} interviewee also supported rejection for formality issues like disclosure \quo{if it's explicitly stated in the call for papers, [as] it was the responsibility of the authors}. 

As rejection is often case-specific, we also discussed how to handle papers with highly promising findings achieved with unethical methods. One \revise{junior} participant argued against rejecting such work, suggesting that it would enable the research community to develop a follow-up solution, because \quo{a lot of brains working together and (...) if that paper got rejected then it's his (authors) own brain to trying to come up with a mitigation strategy}. 
Nonetheless, the general view throughout the interviews was that even highly impactful results should be rejected if obtained through unethical practices.

In the past, some conferences have accepted ethically controversial work but attached a public statement on the first page (e.g., \cite{solomosTalesFaviconsCaches2021, burnettEncoreLightweightMeasurement2015}). Interviewees expressed mixed views on this practice. On the one side, one \revise{senior} participant said, the intention was \quo{to make transparent indeed what was this process and how the decision came to be}. 
Another \revise{PC-Chair} added, it also deals as \quo{a warning that you don't do work like this}. 
On the other hand, \revise{a PC-Chair said,} such a \quo{paper has a bit of a stigma attached to it anyway, as it says right on the first page, which also has a slightly discouraging effect.} 
It \quo{doesn't set a good example}, it would set a precedent, \revise{one senior member noted,}
, especially \revise{another senior added,} since preprints are often shared on \quo{open public servers, either ePrint or arXiv}, 
without the statement, and such warnings are thus overlooked.

The fact that papers may \revise{simply} be published \revise{elsewhere} is also a \revise{common} counterargument when discussing ethically motivated rejections. \revise{A junior and a senior} interviewee \revise{largly} dismissed this concern, noting that such publications have \quo{basically no impact} 
and do not count in academic \quo{bean counting}. 
\revise{The junior} \revise{suggested that} \quo{if the top conferences do that, will Tier 2 conferences eventually do the same, and will it be a kind of trickle-down effect}. 


\takeaway{
The right to reject papers on ethical grounds is widely seen as essential: it prevents the publication of unethical research (Goal \revise{Set} 1), maintains the conference's reputation (\revise{GS} 4), and enforces community standards. It is viewed as the last resort when dialogue fails or harm is irreversible, even for highly innovative results. However, rejection alone raises ethics awareness only for the individual (\revise{GS} 3) but does little to raise it across the broader community, highlighting the need for additional, more transparent interventions.
}

\subsubsection{Research Ethics Committees (REC)}

RECs are the operative bodies executing the in-depth ethical reviewing of papers flagged by reviewers. Consistent with prior work ~\cite{ramuluEthicsSecurityResearch2025}, our participants perceived RECs as helpful to \quo{validate, they review the concerns that people have. And they help trying to clarify these concerns} with \quo{questions back to the authors} \revise{as one senior interviewee put it}. 
\revise{Three} participants \revise{explicitly} emphasized that an explicit REC brings more structure to the review process and helps maintain an overview of ethical aspects in our community. 

Interviewees noted that RECs are a relatively new and still evolving body. Junior members, in particular, reported limited insight into how they operate and called for greater transparency.
Another \revise{concern} was that REC service takes time away from regular PC duties, an issue critical with the community expanding. Interviewees also mentioned that RECs are relatively small compared to full-PC discussions in earlier days, a scalability trade-off. 
While discussing ethical issues in the entire PC seems infeasible, one \revise{PC-Chair called} for \quo{more diversity in terms of backgrounds and geography}. 

\revise{Three} junior participants expressed the wish to consult RECs earlier about ideas, 
\revise{whereas at least four senior members consider} it infeasible given limited resource capacity.
However, according to \revise{two} senior members, there was \quo{a service where you could potentially [go] if you had questions about ethics with your paper}, 
but it was discontinued due to little interest from the community and because 
\quo{cases that require an ethics committee's input might not be ones where the authors [make in-depth] ethical consideration beforehand}. 

Lastly, participants worried that not all papers are reviewed by the REC,
because technically focused reviewers may overlook ethics. Hence, one participant suggested random REC sampling. 

\takeaway{
RECs are widely valued as a ``second set of eyes'' that helps to prevent publications of unethical work (\revise{GS} 1), structure ethical procedures and raise awareness within the community through policies (\revise{GS} 2), and provide ethics feedback to authors (\revise{GS} 3). However, they should improve their transparency and diversity, and expand their scrutiny beyond flagged papers.
}

\subsubsection{(Mandatory) Ethics Sections}
\label{sec:interventions:mandatory}

A section to reflect on the ethical considerations of a study has been standard in some \revise{disciplines}, such as usable security, but USENIX '25 was the first top-four conference to mandate them for all papers. Interviewees viewed these sections as necessary for some papers, but expressed mixed feelings on the mandatory requirement. \revise{According to a PC-Chair,} \quo{it helps and doesn't help at the same time}. 

More than half of the interviewees explicitly mentioned that ethics \revise{sections} help them to reflect on normative aspects of their research.
A key motivation for making them mandatory is seen in the fact that \quo{the papers where [...] we have the most ethical problems are the ones that don't realize that ethics is applicable.} 
To support this change, ethics sections now get an extra page outside the page limit. Interviewees saw it as necessary because some projects require extensive ethical explanations. 

Interviewees viewed the standardization as helpful for both PCs and authors.
For the PCs, \revise{they mentioned that} making researchers' previously implicit considerations explicit to readers helps resolve misunderstandings beforehand, and supports tracking ethical norm develop over time.
For authors, especially juniors, participants noted ethics sections serve as guidance and inspiration for ethics considerations in future studies.
\revise{They also emphasized that the sections only require little effort; for papers with no ethical exposure a short statement explaining why no ethics is involved should suffice.}

Despite the low effort required, \revise{a PC-Chair} noted about the new stipulation: \quo{almost anyone I've talked to hates it}, 
mainly due to the overly descriptive enforcement via desk rejections. 
Another \revise{senior} participant warned that too strict rule enforcement could be counter-productive and may eventually lead to minimal-effort compliance (e.g., via LLMs), similar to how complex password policies led to password reuse and weaker passwords~\cite{inglesant-10-cost-policies, komanduri-11-policy}. 
Others countered that it is lastly the authors' responsibility to read CfPs.

Six interviewees noted that not all work required in-depth discussions of ethics, citing cryptographic studies, local code analysis, or SoKs as examples. 
Some researchers in these areas reported uncertainty about what to write and called for clearer guidance. 
However, USENIX counter-argued that as an application-focused conference, they expect ethics to play a role for every submission~\cite{Guidance84:online}.

\revise{Hesitance} around ethics and the lack of guidance for areas new to these considerations may have led to vague ethics sections.
Two \revise{junior} interviewees mentioned they copy\&paste from other papers, \quo{basically just repeat(ing) the same words again again again} 
, or used LLMs to write this section, reflecting minimal-effort compliance.
One senior interviewee viewed this as a transitional issue, addressing it through review comments rather than a major concern.

Lastly, two participants questioned the cost-benefit balance, \revise{noting} that historically few submissions had raised serious ethical concerns. They \revise{cautioned} not to punish everybody for the wrongdoing of few, 
especially in light of other challenges such as collusion rings~\cite{littman2021collusion}, \revise{which may also require attention}. 

To make ethics sections more structured and to reduce the risk of formal rejections, several participants suggested adopting checklist templates similar to those used by NeurIPS~\cite{NeurIPSChecklist} or ICWSM~\cite{ICWSMChecklist}. Such lists should contain standardized questionnaire items as well as open-ended fields to keep flexibility. 

\takeaway{
Mandatory ethics sections encourage authors to reflect on the ethics of their research (\revise{GS} 3) and, given the extra page, have little cost. Making them mandatory enforces this awareness across the community (\revise{GS} 2).}

\takeawayWO{
At the same time, limited guidance and overly strict enforcement risk vague compliance, resistance, and policy bypasses; a low-barrier checklist template was suggested.
}

\subsubsection{Ethics Framework and Policies}

Lastly, we covered several policy-related interventions: the request to disclose IRB decisions in papers, suggesting coordinated vulnerability disclosure guidelines, using field- or method-specific ethics categories to prompt relevant considerations, and recommending general ethics frameworks.

As for IRBs, \revise{a senior interviewee} pointed out, they \quo{are necessary, but not sufficient.} 
There are many reasons for that, as noticed in literature~\cite{garfinkel2008irbs, ramuluEthicsSecurityResearch2025}. Our interviewees pointed out that IRBs are country- and institution-specific and often lack technical expertise.
\revise{At the same time, two senior interviewees noted,} if authors would submit IRB material alongside their submissions (similar to artifacts), it may reduce later REC requests. Our CfP analysis shows that mentioning IRBs in the paper was often demanded in the past, but shifted, e.g., CCS '26 states that IRB approval is not strictly required and asks authors rather to reason about ethics beyond IRB requirements. This conference seems to follow the recommendation by \citeauthor{ramuluEthicsSecurityResearch2025}~\cite{ramuluEthicsSecurityResearch2025}.

Interviewees agreed that coordinated vulnerability disclosure is crucial and \quo{probably the one area where we [...] have the most experience, and therefore we have kind of the most potential rules}. 
They saw guidelines as especially helpful for new people in the community, also since disclosure is often handled with support from senior group members. 
Nevertheless, some participants emphasized the need for flexibility, citing various reasons for deviating from standards, for example, prior bad experiences with specific vendors that broke embargoes or with hardware vendors with long product and development cycles.

Method-specific ethics categories were viewed by interviewees as useful to prompt authors to very field-specific concerns, noting that the community is too broad for \emph{one} set of rules. 
However, \revise{they said,} categories cannot cover all edge cases and must be updated as new developments emerge, such as AI research. For example, one theorist had difficulties mapping their work to a category and then turned to related work and LLMs for advice. Also, several participants called for example papers alongside categories, noting that \quo{it's good to have at least examples [...] as to what you need to discuss.} 

Regarding general ethics frameworks (e.g., stakeholder analysis), participants have a similar opinion as before. They provide helpful guidance, particularly for authors writing an ethics section for the first time.
One interviewee recalled their USENIX '25 experience: \quo{I thought more about ethics than I did for any other conference} and now uses this reflection as a baseline for future papers. 
Stakeholder analysis was also positively assessed, for helping researchers to better \quo{understand their own research and where their research fits in [the] world better}, eventually improving the paper overall. 

However, as before, participants emphasized the need for flexibility to divert from proposed frameworks. RECs could instead request more structured analysis during rebuttal if needed. They argued that ethical issues vary across research areas, for which some frameworks might fit better than others. Also, the frameworks we use might not reflect the socio-cultural diversity in our community.

\revise{Half of the} interviewees also \revise{noted} \quo{that, like every year for the same conference, they change the guideline.}, which makes it difficult to follow. 
\revise{One junior participant highlighted} that it could increase desk rejects due to minor mismatches and add the burden of rewriting ethics sections for different conference frameworks. 
Many participants therefore emphasized: there should be a community-wide consensus establishing stable and shared ethics standards. 

\takeaway{
Ethics frameworks and policies help to reach goals by outlining ethical expectations for harm-avoiding research (\revise{GS} 1), shaping community norms (\revise{GS} 2), guiding authors and reviewers by giving structure to ethical considerations (\revise{GS} 3), and presenting the community's ethical norms to \revise{the public} (\revise{GS} 4).
However, they should remain flexible in content, yet consistent in process across conferences. 
}

\subsection{Problems And Challenges}
\label{sec:problems}

The analysis of historical developments, existing interventions, and interviewee statements allowed us to identify a number of challenges affecting ethical review procedures.

\textbf{(P1) Growing community}: The growth of the S\&P community increases workload for conference boards, amplifies reviewer fatigue, and consequently slows down PC discussions. 
This constraint hinders the community's ability to approach issues effectively.

\textbf{(P2) Diversity}: As the community becomes more diverse \revise{by expanding into more research fields and geographic regions}, there is a need for RECs to become more diverse and inclusive.

\textbf{(P3) Education}: 
The topic of ethics is often taken for granted, but some members of our community, especially those who joined recently, miss education and guidance on the shared norms.

\textbf{(P4) Transparency}: Transparency at both levels, conference-wide decisions (e.g., rule changes, REC procedures) and review level (e.g., feedback on ethics sections), could be improved, especially for researchers not yet embedded in senior community circles. 

\textbf{(P5) Regulatory Balance}: Conferences need to properly calibrate the scope of formal policies and their enforcement to avoid overregulation and unintended backlash.

\textbf{(P6) Consistency}: \revise{A senior participant noted} inconsistencies in conference procedures: \quo{[E]very PC chair wants to conduct a few experiments. [...] experiments are good, they help us learn more, but we conduct a little too many experiments without evaluating them.} 

\textbf{(P7) REC after research execution}: Interviewees also highlighted the well-known problem~\cite{hantkeWhereAreRed2024, dirksenDontPatchResearcher2024, ramuluEthicsSecurityResearch2025} that RECs assess studies only post hoc, unable to prevent unethical studies in advance.

\subsection{Proposed New Interventions}
\begin{figure}
    \centering
    \includegraphics[width=0.8\linewidth]{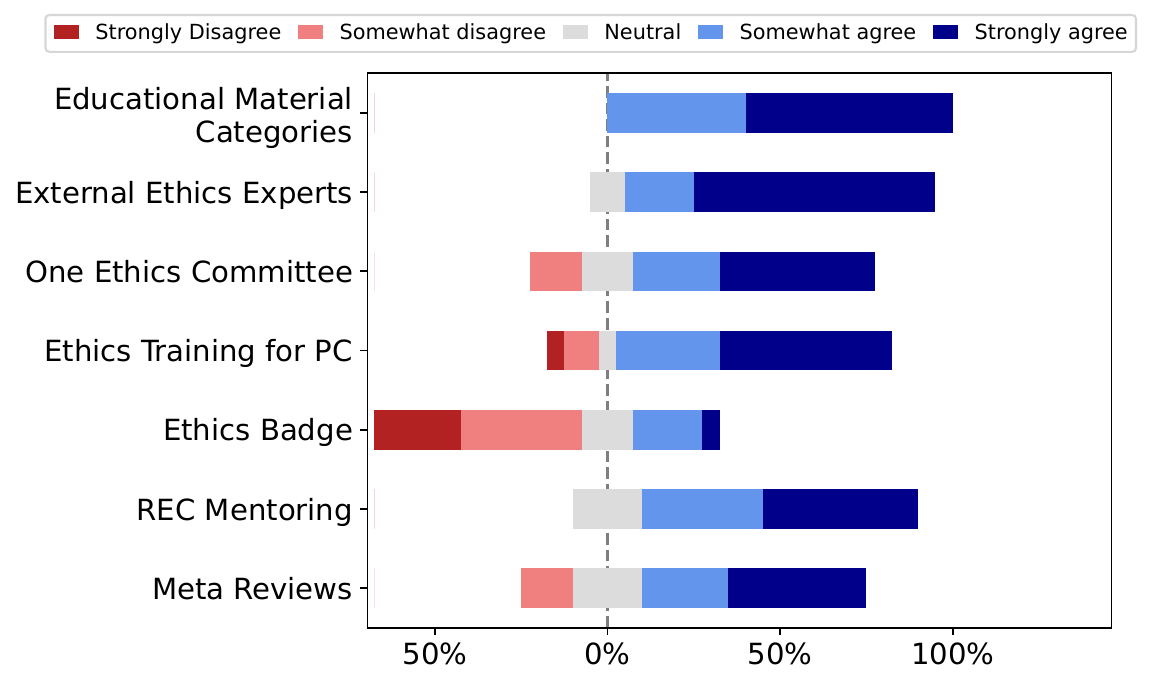}
    \caption{Likert scale of interviewees' opinions about proposed interventions.}
    \label{fig:suggested_likert}
\end{figure}

After identifying four perceived goal sets and key implementation challenges, we now discuss interviewees’ assessments (see also~\Cref{fig:suggested_likert}) of additional ethics interventions that we propose with the aim of addressing problems (\emph{Ps}) summarized in~\Cref{sec:problems}.

\subsubsection{Improvements to Research Ethics Committees}
As a first group of items, we proposed improvements to the REC.

\textbf{Consolidate all RECs into one unified board for all top-four conferences.} Besides unifying ethics reviews, the idea was to have one board that considers and discusses all mentioned \emph{Ps}. \revise{Overall}, our participants liked the idea (average Likert score (LS): 4.00), but saw difficulties in the practical implementation.

\revise{A senior participant} emphasized that \quo{we submit the same papers to the same place [...] We have the same reviewers}. 
and \quo{research ethics is a common problem [...] to solve this problem, we need to be more efficient}. 
One participant even noted that they saw a paper rejected on ethical grounds at one top-four conference but accepted at another. 
Therefore, they would appreciate a unified group that would introduce more consistency and efficiency to ethics reviews. 

However, even supportive interviewees raised concerns regarding its feasibility. First, conferences have tight timelines and \quo{getting enough people to work across multiple conferences is a challenge} \revise{as one PC-Chair put it}. 
But the larger concern, \revise{as another PC-Chair mentioned,} is the institutional fit as \quo{PCs are maybe the same, but the management is quite different}, 
raising questions of funding and accountability.
As a middle ground, interviewees suggested an inter-conference body in charge of general principles and policies, while leaving implementation and case handling to individual conferences.




\textbf{Include external ethics experts in RECs to add perspectives beyond the S\&P community.} This intervention addresses \emph{P2} and was strongly supported (LS: 4.60) with \revise{a junior} interviewee \revise{noting} that \quo{more voices, the better}. 
\revise{A senior viewed it as} \quo{something [...] that brings less load to the people with a very big gain}. 
It would \quo{help existing members of the community understand practices from beyond and incorporate those into our own conception of ethics}. 


Suggested experts included philosophers and sociologists, \quo{people who are actually concerned with ethics, who studied ethics}. 
Some also advocated to include legal experts to advise on data protection, regulatory compliance, and potential legal implications of studies.

Nevertheless, participants also raised some concerns. These include potential community barriers (e.g., terminological differences), a lack of domain knowledge, and an increased coordination overhead that could result from communication with external experts.

Notably, \revise{two} interviewees highlighted that USENIX Security already integrated philosophers into its REC in 2025, pairing them with technical members for joint reviews, an approach reported as effective in bridging domain and communication gaps.

\textbf{Mentoring system for RECs.} Our idea was to pair a junior REC member with a senior one for a joint review during the first year, followed by independent reviews the next year, and mentoring a new member afterward to address \emph{P1} and \emph{P2}.

This idea was positively received (LS: 4.25).
Participants viewed it as a way to diversify the REC as
\quo{more people, including junior staff, may have different perspectives on such matters}, 
and to standardize REC member recruitment, which is currently ad hoc: \quo{we chose people [...] because we know that they [...] have worked previously on some ethics on papers}. 
It is also seen as a low-effort way to spread knowledge of ethical norms within our community.

While some saw mentoring as low overhead, one raised concerns about the time costs required for good mentoring and about long-term sustainability when keeping the same members involved across multiple cycles.
Other concerns included remarks that mentors would bias ethical judgments, 
and skepticism about the added value, as ethics is mainly shaped through life experience.
One senior participant also felt uncomfortable mentoring without formal ethics expertise, but would support it if voluntary.


\subsubsection{Feedback to the authors}
Some participants, especially junior researchers, noted missing feedback (\emph{P4}) on their ethical considerations, motivating our proposals for ethics-focused author feedback.


\textbf{Ethics meta reviews.} We proposed to use meta-reviews as an approach to summarize the PC ethics discussion per paper. Interviewees viewed it positively, but less strongly than other ideas (LS: 3.90), with senior members (LS: 3.70) slightly less supportive than junior members (LS: 4.10).

Supporters of ethics meta-reviews noted that more feedback and transparency are good. \revise{A PC-Chair noted:} \quo{people need some visibility of what was discussed, or otherwise they don't learn}. 
\revise{A junior said, it} could help others see \quo{what was good about this paper, what was wrong with this paper}. 
One \revise{junior} interviewee also noted there is no reason not \quo{to publish these reviews because actually it has already costed the reviewers some time to discuss this part}. 

More skeptical interviewees, mostly senior, questioned whether the benefits justify the added workload given reviewer fatigue, and whether anyone would read meta reviews. 
\revise{One} senior member expressed skepticism regarding the content of such meta reviews: since ethics discussions are often only about asking authors whether they \quo{did or did not address their issue}, there is not much interesting to learn from. Also, these discussions may include sensitive details that should not be public, e.g., if \quo{researchers choose to knowingly violate the law in a minor way [...]. This is a challenging conversation to have, but I don't know if we want a written [statement]}. 



\textbf{``Profound ethical considerations'' badges}. This idea was in analogy to artifacts evaluation badges dealing as feedback (\emph{P4}) for authors and good examples for young researchers (\emph{P3}). However, interviewees were largely skeptical about this idea (LS: 2.45).

A few interviewees saw an ethics badge as a useful way to highlight exemplary ethical considerations for others to learn from and as good feedback \quo{to reward the one[s] that go beyond} the baseline.

However, the effort put into the considerations is not necessarily reflected in the section itself, and ethical complexity varies widely across areas, as multiple participants noted by raising fairness concerns.
The main criticism was that, unlike artifact evaluation, ethics is too subjective to be measured with clear criteria, leading to uncertainty about what the badge would actually stand for. 
\revise{Participants} also argued that ethical considerations constitute a baseline requirement, implying every paper would need the badge.

Instead, \revise{two} interviewees suggested alternative forms of recognition, such as community awards to honor distinguished ethics sections to inspire future researchers with good examples. 

\subsubsection{Education}

Little ethical education (\emph{P3}) beyond general university courses was a recurring concern and may hinder ethical awareness (goal \revise{sets} 2 \& 3) within our growing (\emph{P1}) community. We therefore proposed several education-focused interventions.

\textbf{Method-specific educational material.} Building on USENIX Security’s method-specific tips~\cite{USENIXSecurity252024} (e.g., experiments with live systems; deception studies), we discussed method-specific guidance for researchers. Interviewees strongly supported it (LS: 4.60).

Interviewees described such material as necessary, especially for researchers unfamiliar with ethics. \revise{A senior participant noted:} \quo{I think it is necessary to provide some materials to help them understand.} 
One \revise{junior} valued method-specific categories as \quo{good. If you can give me categories, which can make it easier to think about ethical concerns}, 
and may also help to find some issues that one might not have considered.
Overall, participants stressed that S\&P researchers are no ethics experts and need more education on ethics.

One junior participant \revise{pointed out} that many authors would only consult such material superficially, \revise{noting} \quo{I would not go through the education material before the ethics section.} 
This raises the question of whether the effort to create such material is worth it, given the community's limited resources. 
Interviewees also raised concerns that ethics is subjective and that method-specific categories might be hard to define optimally. Thus, one \revise{senior} participant would prefer to see good example cases rather than fixed categories. 

Multiple interviewees expressed the need for concrete example studies -- rather than abstract scenarios -- as those would ease consideration of ethics.
One of them suggested a database of real studies linked to specific ethical policies, and noted mandatory ethics sections now already create the raw material for this, as we get \quo{more data on what people have thought about, which you can possibly draw on if you find yourself in a similar situation}. 

\textbf{Ethical training for reviewers.} Besides to address the mentioned problems (\emph{P1 \& P3}), this interventions addresses the consistency of decisions (\emph{P6}). This suggestion was received positively but raised concerns about time constraints (LS: 4.10).

Training was seen as a good option to familiarize reviewers with conference-specific ethics policies, as some PC members seem to be unfamiliar with them, and to establish a shared baseline, given that most S\&P researchers lack formal ethics expertise.

One \revise{senior} interviewee, however, noted that reviewers nowadays need ethical expertise mainly in their own subfield, with which they should already be familiar.
\revise{A junior} participant was afraid that training could become overly rigid and \quo{brainwash} trainees to adhere to one specific ethical approach, constraining reviewers' independent judgment; teaching general ethical ideas was seen as acceptable.
Others asked who would provide the training, particularly against the background of different ethical traditions and socio-cultural backgrounds. 
And even with sufficient training, some interviewees wondered whether it would pay off, since \quo{reviewers who don't care about this will [...] just skip over this as fast as possible.}, while those who care are already knowledgeable about it. 
Hence, \quo{there's a difference between providing the training and ensuring that people take it} 
Above all, the main concern was the severe time constraints and reviewer fatigue: \quo{we already scramble, we already struggle a lot to get enough reviewers [...] it's a mess.} 


\revise{It was} pointed out that many researchers already receive ethics training from IRBs. \revise{One junior suggested} \quo{if the reviewer has participated in the training [...] they should not need to participate in the same training for conference B in the same year}. 
Lightweight briefings on updated ethics policies were widely appreciated.

\takeaway{
\revise{We proposed a range of new interventions, with improvements to RECs and education/training receiving positive support, while feedback-related ideas were seen as the least important among the proposed ideas. Interventions were favored when they offered clear benefits with low overhead, whereas more ambitious proposals (e.g., centralized RECs) raised feasibility concerns, highlighting the need for lightweight solutions that respect people's workloads.}
}

\subsection{Review Form}

\begin{figure}
    \centering
    \includegraphics[width=0.8\linewidth]{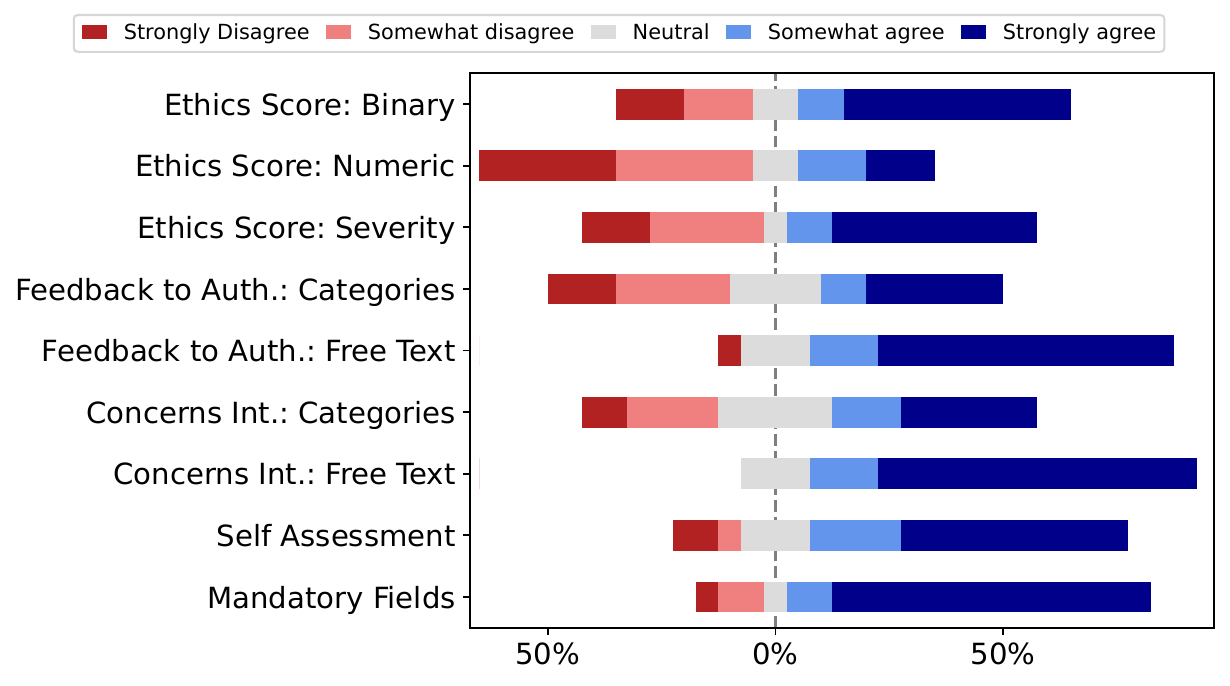}
    \caption{Likert scale of interviewees' opinions about items in review forms.}
    \label{fig:template_likert}
\end{figure}

As outlined before, reviewers serve as the primary agents of ethical scrutiny in the publication process. RECs become involved only when reviewers identify potential ethical concerns and flag them in the review forms. Consequently, the inclusion of well-designed ethics-related items within these templates represents a significant ethics intervention in its own. However, one senior interviewee stated, \quo{the USENIX Security 26 review form is too complicated. [...] that's feedback I've heard from various people.} 
Against this backdrop, we discussed with our interviewees several approaches to communicating ethics-related considerations via review forms. We present the identified findings below and in~\Cref{fig:template_likert}. 

\subsubsection{Ethics Rating}
We discussed three ways to rate or flag ethical concerns in reviews: a \textbf{binary flag}, a \textbf{numeric scale}, or a \textbf{severity score} (e.g., \emph{no issue}, \emph{minor}, \emph{severe}).
Interviewees were divided on the use of scores, but agreed on two points: (1) numeric scores are the least useful (LS: 2.55) since they offer little added information beyond severity; (2) free-text justification along scores is essential.

Severity score (LS: 3.45) was valued by some \revise{senior members} as a way, \quo{to give [...] the chairs some guidance} and help them to prioritize cases. 
Also, it was noted that ethics is too nuanced for a mere yes or no.
On the other hand, others preferred a binary flag (LS: 3.65), arguing that a paper either raises ethics concerns or does not.
Three interviewees rejected scores altogether as they make reviewers think less about ethics. \revise{One PC-Chair highlighted} that
\quo{what is important is, that people tell that there is a potential issue and explain why and that then other people have a look at it}. 

\subsubsection{Ethics Concerns}
Regarding communicating concerns, we discussed \textbf{PC-internal feedback} and \textbf{feedback to the authors} in the form of either \textbf{free text} or as \textbf{pre-formulated categories}. 

Across both channels, interviewees \revise{strongly} preferred free text (LS: authors 4.35; internal 4.55).
Categories (LS: authors 3.15; internal 3.35) were seen as a supportive instrument, but not sufficient on their own.
Although some ethics problems occur frequently (e.g., missing information about a CVD process) a lot of ethical feedback is paper-specific and pre-formulated categories cannot cover all cases.
Also, regarding feedback to authors, one \revise{junior} interviewee emphasized that every rejected submission deserves that we
\quo{make the effort to write a few sentences [...] explaining why we don't want this.} 
For internal feedback, some senior members pointed out that it helps the REC to internally assign papers to REC members with corresponding expertise.
Also, \revise{a PC-Chair} noted that internal feedback is important, because sometimes, \quo{you'd want to raise to the committee that you wouldn't necessarily want to raise to the authors, because you're unclear whether something constitutes violations} 

\subsubsection{Additional Proposals}

Lastly, we proposed a \textbf{self-assessment} score for reviewers' ethics \revise{confidence} and the requirement to make ethics fields in reviews \textbf{mandatory}.

The self-assessment was viewed as moderately useful (LS: 3.95), mostly for helping RECs to prioritize and double-check papers with low \revise{reviewer} confidence.
However, \revise{a junior member raised concerns} that it could lead to situations where reviewers or authors ignore valid concerns expressed by less confident members.
\revise{A senior} participant also noted that RECs should review every flagged paper regardless of a reviewer's self-assessment.
%
Most interviewees supported mandatory ethics fields in reviews (LS: 4.30).
Critics mainly argued that not all papers have ethical aspects, but were comfortable with mandatory binary flags or scores as long as they can indicate \emph{no concern}.

\takeaway{
\revise{Review forms should remain as simple as possible to avoid overburdening reviewers while still enabling ethical scrutiny. While binary flags or severity scores serve as a good first indicator, free-text explanations are essential. If pre-formulated answer categories are used, they should only serve as support. Fields should be mandatory, with the option to indicate that no concerns exist.}
}
\section{Discussion}
\label{sec:discussion}

One can reason about good or bad practices only when knowing intended goals. We therefore identified the perceived goals of research ethics in the S\&P community and use them as a baseline to discuss interventions, encouraging decision-makers to do the same.

\subsection{Goals of Ethics Interventions}
According to our interview analysis, members of the S\&P research community perceive four sets of goals behind ethics interventions at top-four conferences (\Cref{sec:goals}). The examination of the existing interventions (\Cref{sec:interventions:existing}) suggests that normative regulation of external impacts of S\&P research (goal \revise{set} 1) as well as external perception and evaluation of the S\&P research community (goal \revise{set} 4) have been already addressed by such institutional measures like ethical requirements written into CfPs, ethical review integrated into the routine reviewing process and supplemented by reviewing activities of specialized bodies like RECs, and ultimately by the right to reject publication of unethical work. This fact can prompt some community members to think there is little left to do regarding the institutionalization of research ethics. However, many interviewees indicated that a further increase in ethical awareness within the S\&P research community and support of ethically sound practices of its individual members constitute another major set of goals (2 and 3) supposed to be addressed by ethics interventions.

We see these four sets of goals as interconnected. Reputation and assessment of the S\&P research community by external actors (goal \revise{set} 4) depends to a large extent on the ability to effectively regulate in normative (ethical) terms various impacts of research activities and results on the outside world (goal \revise{set} 1). And the approximation of the latter goal is significantly influenced by the community’s ethical awareness and the ability of its individual members to put the principles of ethically sound science into practice. Against this background, we propose \revise{below} an expansion of ethics interventions specifically aiming at a further rise of ethical awareness within the S\&P research community and making it easier for individual community members to integrate ethical reflection into all stages of their scientific work.

\subsection{Ethics Interventions Recommendations}

\revise{As mentioned earlier, the community has extended the scope of ethics interventions since the early 2010s. Still, we identified a few improvements the S\&P community should discuss and consider how to put them into action.}

\textbf{Rejection is the ultima ratio and needs \revise{careful} consideration.}
Rejection stands above all other interventions, which either inform or operationalize this decision. It is largely accepted and uncontroversial \revise{among our interviewees}. Disagreement arises only from ethically ambiguous research with great results, for which some have decided to publish with an ethical disclaimer. The impact of this approach remains unclear~\cite{ramuluEthicsSecurityResearch2025}.
Irrespective of whether the work is accepted, the harm to the outside world (goal \revise{set} 1) has already occurred in such cases. 
Given the lack of a community-wide standard for ethics, rejection may lead to unethical \revise{work being published} at another conference, causing negative impact on the community (goal \revise{set} 2). Similarly, not publishing the work may lead to other researchers conducting the same or similar experiments, causing additional harm to the outside world. However, publishing with a note may suggest that transgressions of ethics are acceptable as long as the results are sufficiently interesting, leading to negative incentives (goal \revise{set} 3) for researchers and potential reputational damage to the conference (goal \revise{set} 4). More positively seen, publication with disclaimers sets a precedent for what should not have been allowed to be published, making it an explicit signal for future authors. We believe this requires community-wide discussions.


\textbf{Research Ethics Committees can improve with greater diversity, more transparency, and better coverage.}
RECs already \revise{contribute to} structuring ethical review procedures (goal \revise{set} 2) and providing feedback to authors (goal \revise{set} 3). However, we \revise{identified} concerns about limited transparency and diversity, and that unethical work may go unnoticed when reviewers focusing on technical aspects miss to flag it. \revise{Including additional} perspectives and greater coverage could help reduce blind spots in identifying harmful work (goal \revise{set} 1), while increased transparency would strengthen the acceptance of norm enforcements within the community (goal \revise{set} 2). To improve, we suggest REC mentoring, including external experts, and, if full coverage is not feasible, \revise{considering approaches such as} checking random samples.

\textbf{Ethical policies should guide practice without becoming over-restrictive.}
\revise{Our interviewees} liked policies as guidance, and we found indicators that example policies (e.g., USENIX) are reused as baseline for future work (goal \revise{set} 2). At the same time, we should be careful with overly restrictive rules, as people called for flexibility. We otherwise risk negative effects and pushbacks (goal \revise{set} 2), such as minimal-effort compliance or policy circumventions (see complex password policies). We encourage decision-makers to open their policies to community input, as recently done by CCS '26~\cite{CCSGitHub}.

\textbf{Ethics section should be mandatory, with clearly defined exceptions.}
While mandatory ethics sections remain controversial as a restrictive intervention, they seem logical towards the awareness goal \revise{sets} (2 \& 3) and typically require little effort. While the community still discusses whether every paper needs one, we argue that a structured, uniform approach has benefits and thus recommend it for every paper.
For example, the USENIX '26 Ethics Guidelines provide four key questions (stakeholders, impacts, mitigations, decision to proceed) that could serve as a lightweight checklist for authors to self-certify whether they should be exempted from discussing ethical implications. This way, all researchers have to reflect and self-assess their ethical implications before submission.

\textbf{The community needs shared, community-wide ethics policies and educational material.}
While CfPs provide valuable guidance~\cite{ramuluEthicsSecurityResearch2025}, frequently changing, conference-specific rules lead to frustration in the community and potential increased reviewer fatigue.
We thus recommend developing community-wide standards (goal \revise{sets} 2 \& 3) that can be adopted across the top four and other conferences (trickle-down effect). Beyond policies, shared resources should include research-area-specific educational materials, concrete example papers, as researchers often learn community-specific ethical understandings from studying related work, and suggested~\cite{hantkeWhereAreRed2024, ramuluEthicsSecurityResearch2025} anonymized REC rejection decisions as a way to communicate community boundaries.

\textbf{The community would benefit from a coordinated ethics steering body.}
Conferences established RECs 
to guide ethical decisions. While conferences need these to take on conference-specific decisions, we advocate for a community-wide ethics steering body to support the above-mentioned shared policies and coordination. Such a body could better scale with the growing community, address transparency concerns, advise on ethically complex research at the design stage, and help follow the identified goals.


\section{Call to Action: An Ethics Wiki}

In our discussion, we advocate for a community-wide effort to discuss and coordinate ethics policies and interventions across all top-four conferences. Yet, the question remains: who should orchestrate this effort? While this paper proposes several recommendations, we believe progress requires a different approach than producing papers. It requires open and transparent discussions within the community to develop an actionable outcome.

To begin this process, we opened a community ethics wiki filled with existing policies and enabling discussions about new developments via GitHub issues~\cite{ethicswiki}. As a call to action, we invite the community, especially those who proposed other promising ideas, to discussions, contribute, and advance the shared ethics standard.

\section*{Acknowledgments}
We thank the CCS reviewers and shepherd for their valuable feedback, which helped improve the paper's structure and flow.

This work was conducted in the scope of a dissertation at the Saarbrücken Graduate School of Computer Science

\bibliographystyle{ACM-Reference-Format}
\bibliography{reference}

\appendix

\section{Generative AI Usage}

In the process of writing the present paper, generative AI tools were used exclusively for proofreading and text shortening.
We had distributed the initial drafting of sections between the authors. Only after these section drafts were completed, the authors were allowed to use ChatGPT and Grammarly to assist with proofreading and shortening text sections. No suggested AI-generated formulations were directly adopted. Instead, suggestions were used solely as inspiration for restructuring sentences, with all final wording written by the authors themselves.
\section{Ethical Considerations}
\label{appx:ethics}

We followed the ethical practices established in the S\&P research community. The present study was approved by our university's Ethical Review Board. Before interviews, all participant were sent the consent form informing them about the content of the interview and their rights as participants. They all attended the interview voluntarily and were explicitly asked before the interview if they have consent-related questions or concerns. They had the option to discontinue their participation in our study at any time and to ask for deletion of their data. \revise{We also deleted all interview recordings (video and audio) after transcription and only kept the textual data.} Besides these common interview-related ethical practices, we had two main challenges in mind that we considered when planning the present study:

\textbf{How can participant anonymity be ensured when the number of potential participants is small?}
Since the participant pool of PC chairs is very small, it would be quite easy to de-anonymize individuals if not taken extra care about this aspect. We therefore decided not to publish demographic data about individual interviewees and also abstained from indicating the interviewee number for verbatim quotations and paraphrased transcript references. Furthermore, we have slightly modified some direct quotations to preserve their meaning while preventing the identification of interviewees.

\textbf{How can participant recruitment within the community be conducted at a sufficient scale while minimizing unsolicited bulk mails, but also not conflicting with too many program-committee members?}
Another challenge was the recruitment of participants within the S\&P community. A broad recruitment strategy via mass emails was not feasible, as it could have interfered with the double-blind review process. At the same time, repeatedly contacting and reminding a small group of individuals until they agree to participate would have been ethically unacceptable.

To get more perspectives on this problem, we first talked to former PC chairs about how they would handle papers that would conflict with too many PC members. Based on this input, we then decided to adopt a targeted and staged recruitment strategy (as described in~\Cref{sec:methods}). Senior members of the community were hand-picked based on their service track-record and contacted via targeted personalized e-mails. Junior members were randomly sampled based on their research backgrounds and contacted then in small batches. In both cases, we sent at most one reminder after approximately one week whenever no earlier response arrived. If no response at all was received, we moved on to a new set of candidates until our sample plan was met. During submission, we asked all participants who were part of the PC whether they would be fine with being marked as \emph{other conflict} in HotCRP, explaining to them that it would potentially leak the fact of their participation to the PC chairs. As an alternative, we proposed that they can self-conflict with our paper. All affected participants were fine with the first option. We believe, in conclusion, this process holds a reasonable balance between minimizing unsolicited contact, avoiding conflicts of interest, and ensuring sufficient sample diversity.
\section{Open Science}
To uphold open science, we make the following artifacts public.

\begin{itemize}
    \item The interview guide and survey instrument we used in our qualitative interviews (\Cref{appx:interview-survey} and \ref{appx:interview-guideline}).
    \item \revise{The consent form and mails used for recruitment (\Cref{appx:consent} and \Cref{appx:recruitment})}
    \item The codebook used for the 
    \revise{qualitative} interview analysis (\Cref{appx:interview-codebook}) and process data analysis (\Cref{appx:process-data-codebook}).
    \item The resulting data from the survey instrument (see~\cite{supplementary}).
    \item The statistical data and scripts behind our keyword analysis (see~\cite{supplementary}).
\end{itemize}


We decided to not publish the interview transcripts due to de-identification risks (see~\Cref{appx:ethics}).
\section{Interview Participants}
\label{appx:participants}

Earlier, in \Cref{tab:interviewees-body}, we present our interview participants, including their community service roles (where applicable), the duration of each interview, and whether they indicated in the pre-questionnaire that they had previously received any ethics training \revise{(both coursework and IRB/ERB training)}. Due to the small set of potential participants, we decided to present demographics aggregated in \Cref{tab:research-locations} and \Cref{tab:research-areas}.

For the research areas, we automatically went through all available HotCRP instances of the major conferences from 2020 to 2024 and queried the unauthenticated \texttt{/api/searchcompletion} endpoint to retrieve the complete list of topics available at each conference. Based on this list, we manually merged semantically similar topics and removed redundancies to compile a holistic list of research fields. While this list is certainly not perfect, we found it to provide a sufficiently comprehensive and appropriately abstracted representation of the research landscape for the purposes of our analysis.

\revise{The research field \textit{System Security} was added by participant I-16 via an open answer field in our pre-interview questionnaire and is therefore listed in the research fields.}

\begin{table*}
    \centering

\begin{tabular}{l | c c c c c c c c}
    Location & US & Germany & Switzerland & Netherlands & Portugal & China & France & Austria \\ \midrule
    Count    & 11 & 3 & 1 & 1 & 1 & 1 & 1 & 1 \\
    \bottomrule
\end{tabular}
    \caption{Interviewees' institute locations.}
    \label{tab:research-locations}
\end{table*}

\begin{table}
    \centering
    \begin{tabularx}{\linewidth}{X l}
    Area Name & Interviewees \\ \midrule
    Operating Systems and Software Security & 8 \\
    Mobile Systems Security & 8 \\
    Usable Security and Privacy & 6 \\
    Applied Cryptography and Cryptographic Analysis & 5 \\
    Machine Learning and Security & 5 \\
    Hardware, Cyber Physical, Embedded and IoT Systems Security & 5 \\
    Privacy and Anonymity & 4 \\
    Research on Surveillance, Censorship and Other Social Issues & 3 \\
    Malware Research & 3 \\
    Cyber-Crime Defense, Forensics and Diagnostics for Security & 3 \\
    Law, Policy, and Ethics of Security & 2 \\
    Formal Methods and Language-Based Security & 2 \\
    Network and Protocol Security & 2 \\
    Cyber Attack (e.g., APTs, Botnets, DDoS) Research & 2 \\
    Secure Computer Architectures & 2 \\
    Trustworthy Computing & 2 \\
    Web Security and Privacy & 2 \\
    Authentication, Access Control, and Authorization & 1 \\
    Cloud Computing Security & 1 \\
    Blockchain and Distributed Systems & 1 \\
    \emph{System Security} & 1 \\
    Wireless Security & 0 \\
    \bottomrule
\end{tabularx}
    \caption{Interviewees' research areas. \textit{System Security} was added by participant I-16 via an open answer field in our pre-interview questionnaire.}
    \label{tab:research-areas}
\end{table}

\section{Keyword Analysis}
\label{appx:keyword}

To understand how the topic of research ethics has evolved over time within the S\&P research community, we conducted a keyword analysis of security and privacy papers published at top-four conferences during the period 2000-2024. \revise{As described in \Cref{sec:methods:keywords}}, in contrast to previous work by \citeauthor{zhangEthicsSecurityResearch2022}~\cite{zhangEthicsSecurityResearch2022}, statistical features such as TF-IDF weighting and word co-occurrences were used to ensure the robustness of the extracted keyword list. Our results indicate an overall increase in awareness of ethical issues, particularly within the botnet sub-community.

\begin{figure}
    \centering
    \includegraphics[width=0.9\linewidth]{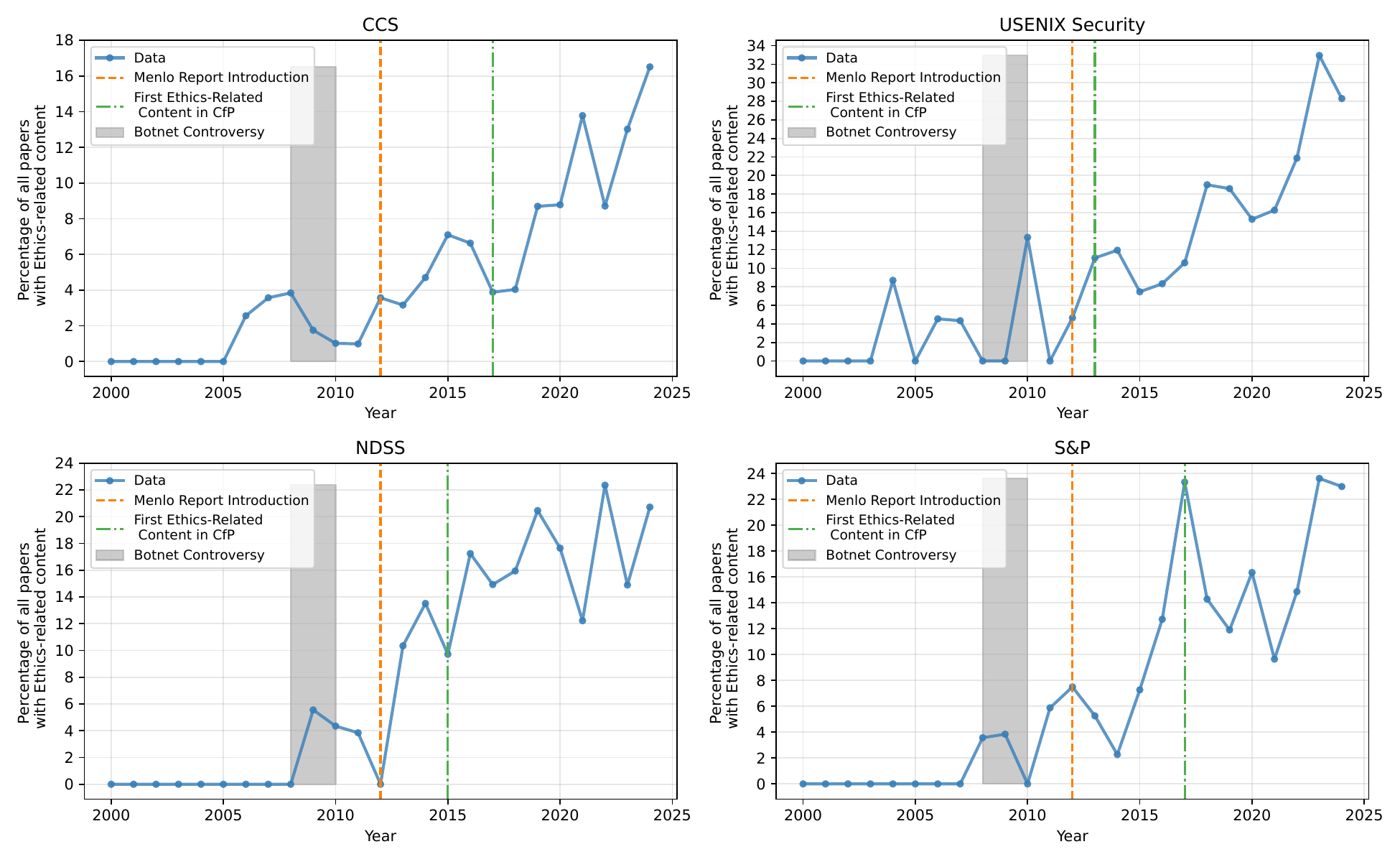}
    \caption{Paper with ethics-related content over time, broken down by conference.}
    \label{fig:general_ethics_over_time}
\end{figure}

\begin{figure}
    \centering
    \includegraphics[width=0.9\linewidth]{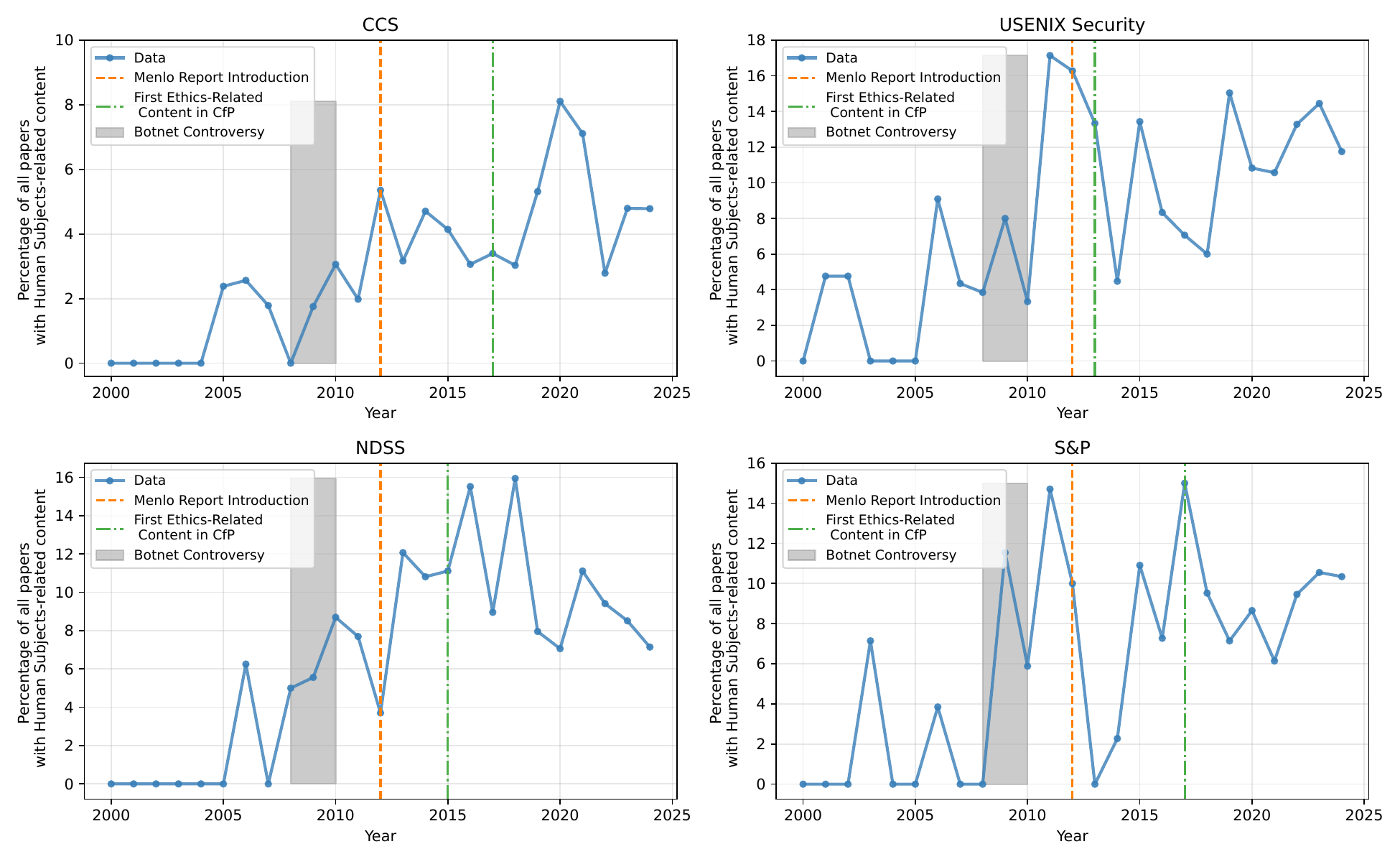}
    \caption{Paper with human subjects-related content over time, broken down by conference.}
    \label{fig:human_subjects_over_time}
\end{figure}

\begin{figure}
    \centering
    \includegraphics[width=0.9\linewidth]{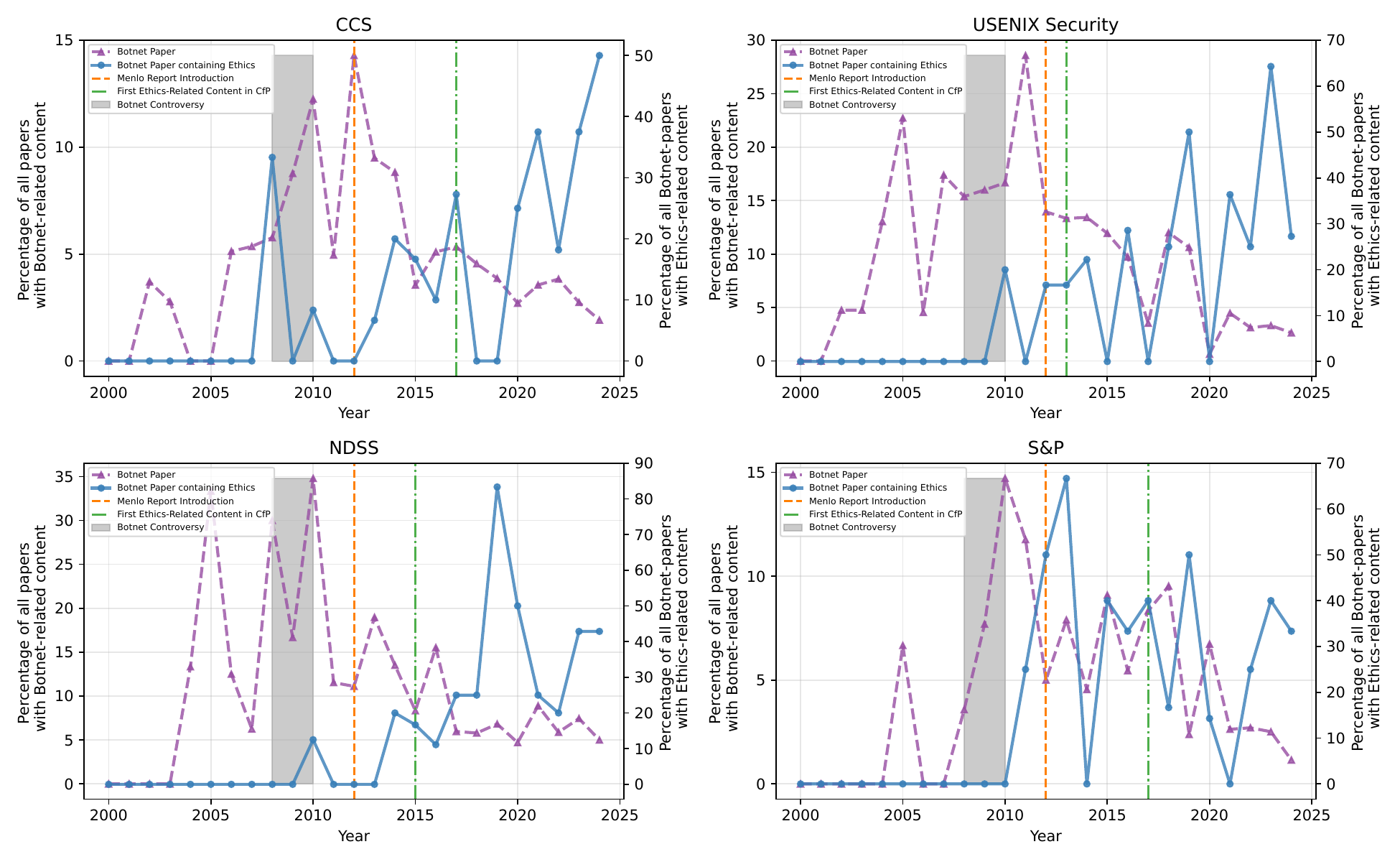}
    \caption{Paper with botnet-related content (dashed line), as well as the share of all botnet-related papers dealing with ethical topics (solid line), over time, broken down by conference.}
    \label{fig:botnet_over_time}
\end{figure}

The results are displayed in the figures below (\Cref{fig:general_ethics_over_time} - \ref{fig:botnet_over_time}). In addition to the data itself, the graphs also show historically interesting points that may have influenced the treatment of ethical issues in S\&P papers. These include the introduction of the Menlo Report, the first mention of ethics-related content in the CfP, and what we have termed the \emph{botnet controversy} described above.

\Cref{fig:general_ethics_over_time} visualizes the data of the \emph{General Ethics} dimension. Across all four conferences, a noticeable, albeit non-monotonous, increase can be seen within the studied period. Since 2012, at least one paper with ethical content has been published at each of the conferences examined. However, there is no clear correlation between the trend and the historical points marked. There are both upward and downward trends during the botnet controversy. With regard to the Menlo Report and the first mention of ethical content in the CfP, no clear trends can be identified in the following years.

The data of the \emph{Human Subjects} topic is displayed in \Cref{fig:human_subjects_over_time}. Similar to the \emph{General Ethics} dimension, a higher presence of the topic can also be seen here in recent years. However, \emph{General Ethics} and \emph{Human Subjects} did not develop in the same way. There were noticeable sharp declines at USENIX 2014 and S\&P 2013. No overarching patterns can be identified with regard to the historical markers either.

In addition to these general ethical categories, we also took a closer look at papers dealing with the topic of botnets (\Cref{fig:botnet_over_time}). With the exception of a slump in 2011 at the NDSS, the number of botnet papers across all conferences rose steadily during the botnet controversy. Around 2012, interest in the topic began to decline sharply, with the proportion of papers containing botnet-related content fluctuating around ten percent in recent years. The NDSS is once again an exception here. Here, the presence of the topic is at a slightly higher level, between ten and 20 percent. However, while the overall presence of the topic has declined since the years following the botnet controversy, an almost opposite dynamic can be observed in botnet papers that deal with ethics-related content. With the exception of the CCS, ethics played hardly any role in the botnet field at the end of the 2000s, but in the years that followed, there were several years in which ethics were widely discussed within this sub-community, such as at CCS 2024, USENIX 2023, and NDSS 2019. This points to an increased ethical awareness in this sub-community.

\begin{table}
    \centering
    \begin{tabularx}{\linewidth}{llX}
     & \makecell{Heading\\Keyword} & \makecell{Keyword\\List} \\ \midrule
    \makecell[tl]{General\\Ethics} & ethic & review~board, institutional~review, board~irb, irb~approval, board~erb, ethical~considerations, ethics~committee, ethical~concerns, ethical~issues, institution~irb, ethical~implications, ethical~principles, ethical~frameworks, menlo~report, belmont~report, public~interest \\
    \makecell[tl]{Human\\Subject} & ethic & consent form, explicit consent, human subjects, real users, involve human \\
    Botnet & botnet & botnet, infected machines, infected hosts \\
    \end{tabularx}
    \caption{Keyword lists for the topics examined}
    \label{tab:keywords}
\end{table}

\clearpage
\onecolumn

\clearpage

\section{Recruitment Mails}
\label{appx:recruitment}

\subsection{For Senior Members}
\label{appx:recruitment:senior}

\noindent\textit{Research Ethics Interview Invitation}

\begin{quote}
\setlength{\parskip}{0.5em}
\setlength{\parindent}{0pt}

Dear <name>,

I’m <author>, a PhD student in <PI>’s group.

We’re currently doing a study on research ethics practices and interventions implemented at our top-tier conferences. To get a better understanding about the current developments and improvements in our community, we’re interviewing current and past PC-Chairs about this topic. Since you are the <position> for <conference> this and next year, we’d love to hear your views.

The interview will be with me and likely another colleague of our research team on Zoom. It typically takes 75 minutes, or 45 minutes in a shorter version.

Let me know if you’re interested and if so, when you’d be free so I can schedule a call. 

Thank you, I’m looking forward to talking with you.

Best regards,\\
<author>
\end{quote}

\subsection{For Junior Members}
\label{appx:recruitment:junior}

\noindent\textit{Research Ethics Interview Invitation}

\begin{quote}
\setlength{\parskip}{0.5em}
\setlength{\parindent}{0pt}

Dear <name>,

I’m <author>, a PhD student at <institute>, and I’m conducting research on how ethics is considered and supported at top-tier security and privacy conferences.

As part of this work, we’re interviewing (randomly sampled) authors, PC chairs, and other contributors to understand how current ethics practices at conferences are received and how ethics in our community changes. Since you recently published at <conference>, your perspective as an author is especially valuable to us. Even if you feel you don’t have much to say about ethics in research, that’s important too; we want to capture a wide range of experiences and viewpoints.

The conversation will take place over Zoom with me and (likely) another member of our research team. It usually takes about 75 minutes (you can also choose a shorter version depending on your availability).

If you’re interested, just reply with a few time slots (together with your time zone) that work for you, and we’ll schedule the call at your convenience. Thank you in advance!

Your input will help us better understand and support/improve ethical practices in our community and at our conferences.

Best regards,\\
<author>
\end{quote}

\section{Consent Form}
\label{appx:consent}

{\parindent0pt
\textbf{Study Name}: Strengthening Ethics Review Procedures

The goal of this study is to get a deeper understanding of the ethics procedures at security and privacy conferences and identify goals, challenges, and potential improvements.

\textbf{Authors}: <list of authors>

\textbf{Data Protection Officer}: <e-mail>

By completing this survey and participating in the interview, you consent to the following:
\begin{itemize}
\item \textbf{Recording}: Your interview will be recorded via Zoom only for research purposes. During one part of the interview, we will ask you to complete a task while sharing your screen. If this is not possible for you, as an alternative, the interviewer can share their screen and be guided by you through the task. The recordings will not be shared with anyone outside our research team. After completing the anonymized transcription of the interview, the recording of it will be deleted.

\item \textbf{Data Usage}: The content of this questionnaire and the subsequent interview will be analyzed only by our research team using standard research methods such as qualitative coding. Its processing is based on your consent in accordance with Art. 6(1)(a) GDPR.

\item \textbf{Data Storage}: Parts of this study are conducted using Qualtrics questionnaire software. Questionnaire data are stored on Qualtrics servers in the EU and are accessible via password-protected user accounts. All data downloaded from Qualtrics servers and recorded via Zoom will be stored in a secure cloud storage system of our institute. All anonymized data and study documents can be kept for a period of at least 10 years.

\item \textbf{Anonymized Quotations}: We may use anonymized verbatim quotes from your interview in research outputs (e.g., publications, presentations).

\item \textbf{Voluntary Participation}: Your participation in this questionnaire and interview is entirely voluntary.
\end{itemize}

If you have any concerns or specific conditions regarding the points mentioned above, please let us know in the space provided below:

[FREE TEXT FIELD]

You have the right to information, correction, deletion, restriction of processing, data portability, as well as a right of appeal to a supervisory authority of your choice.

Your participation is voluntary. If you decide to participate, you can cancel your participation at any time and revoke this declaration of consent with effect for the future. In this case, we will also delete all data collected from you until then. Please note that after completion of the study, all data will be anonymized and therefore deletion of your data is no longer possible from this point on.

If you decide to end your participation, if you have any questions, concerns, or complaints, or if you wish to report a violation related to the study, please contact the person(s) responsible for this study or the responsible data protection officer.

This study was reviewed and approved by <institute> Ethics Review Board in accordance with the guidelines for research involving human subjects.

[] Yes, I consent to the above, begin the study.

[] No, I do not consent, I do not wish to participate.
}

\section{Interview Guide}
\label{appx:interview-guideline}

The interview guideline (\Cref{tab:interview-guideline}) asks about the participants' experience with existing ethics procedures at academic conferences, their view on specific ethics interventions, the reasoning and goals behind the implementation of these practices, and their assessment of the importance of ethics in research. This guideline exists in slightly modified versions for PC-Chairs with and without a REC, general senior people, and authors who have never been part of a PC. For the latter, we rather asked if they ever got feedback regarding ethics or got otherwise in contact with ethics at a conference. 

\begin{table}[h]
    \centering
    \begin{tabularx}{0.9\textwidth}{l X}
        \toprule
        Introduction &
        Before we dive in, how would you describe your overall experience with ethics-procedures at academic conferences as a PC chair or member and as an author? \\ \midrule

        Ethics Procedures &
        Let's start right into the interview: You are/were the chair of CONFERENCE. CONFERENCE has/had a research ethics committee supporting the review phase. Could you describe this procedure for a submitted paper? \\ \midrule
        
        Ethics Interventions & We explain our definition of research ethics intervention: Everything at a conference that influences how we consider ethics in our research, e.g., ethics reviews, guidelines, other information provided. \\
        & We will go through some ethics interventions that conferences had implemented in the past. Please tell me your opinion about each, is it helpful or not? 
        
        \begin{itemize}
            \item Research Ethics Committees: S\&P and USENIX have a REC since 2022, NDSS followed 2025, what do you think about them?
            \item  Mandatory Ethics Section: USENIX made ethics sections mandatory for all paper in 2025. What do you think about this?
            \item IRB Decision Disclosure: All conferences ask for information about IRB decision for papers. What do you think about this?
            \item Responsible Disclosure Requirements: S\&P and USENIX explicitly recommend how to conduct responsible disclosure. What do you think about this?
            \item Ethics Categories: Some conferences list what kind of research and methods require ethical considerations. For example, human subject research, research on live practices, or well-being of the research team. What do you think about this (ethical categorization in general)?
            \item Right to reject: Some conferences mention explicitly that they might reject unethical paper. What do you think about this?
            \item Ethics Frameworks: Conferences suggested various frameworks to evaluate ethics in the past. For example, evaluating harm vs. benefit or a stakeholder analysis. Are such frameworks helpful or should authors be free in deciding how they discuss ethics in their papers?
        \end{itemize}
        \\
        & Do you know of any other interventions implemented to ensure ethical considerations during paper writing and reviewing at the conference? Please explain the details of the intervention to me. \\ \midrule

        Survey Tool & In this part of the interview, we let the interviewees share their screen and go through a survey tool (see~\Cref{appx:interview-survey}). \\ \midrule

        Reasoning &
        What was the reasoning behind the decision to implement ethics practices and interventions? \\
        & Were there discussions or debates in the PC about this implementation and alternatives? If so, what were the main arguments? \\
        & Have you or your predecessors tried out other instruments or frameworks? If so, which ones, and how did they perform? \\
        & Did the PC consider insights from other fields (e.g., sociology, philosophy, medical research, psychology) to design these interventions? \\ \midrule

        Goal &
        In your view, what are the primary goals of implementing research ethics practices at conferences? \\
        & Are these goals already reached, or is it an ongoing process towards these goals? \\
        & (Optional) How well do you think the current approach achieves these goals? \\
        & What could be improved with the current practice? \\
        & Are there any aspects of the current ethics practices at conferences that you don’t like? \\
        & What are your thoughts on the development of ethics considerations in our community? \\
        & (Optional) Do we need IRB approvals and committees at conferences at the same time? \\ \midrule

    Importance of Ethics & 
    What are your thoughts on how to handle unethical papers that have a great contribution? \\
    & Stepping away from your role as PC chair, how do you personally see the importance of ethics in research compared to other aspects, such as novelty or technical correctness? \\ \midrule

    Ending &
    Is there anything else you'd like to add about the ethics review procedure? \\
        \bottomrule
    \end{tabularx}
    \caption{Interview Guideline}
    \label{tab:interview-guideline}
\end{table}

\section{Survey Tool}
\label{appx:interview-survey}

\begin{figure}
    \centering
    \includegraphics[width=.8\linewidth]{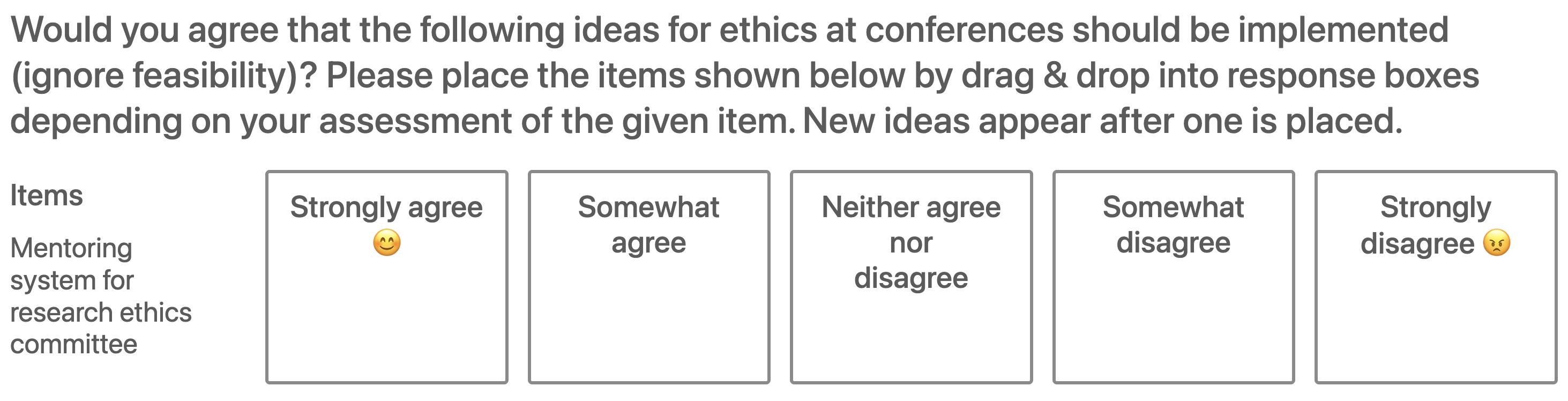}
    \caption{Survey instrument used as part of the interviews.}
    \label{fig:survey}
\end{figure}

As one part of the interview, we let the interviewees share their screens and go through a survey tool. The tool shows several suggested interventions in the form of items that the participant drags and drops in buckets labeled with a Likert scale (see~\Cref{fig:survey}): Strongly agree -- Somewhat agree -- Neither agree nor disagree -- Somewhat disagree -- Strongly disagree. The introduction text and items of the tool are listed below.

\itembox{
Would you agree that the following ideas for ethics at conferences should be implemented (ignore feasibility)? Please place the items shown below by drag \& drop into the response boxes depending on your assessment of the given item. New ideas appear after one is placed.
}

\begin{itemize}
    \item Publishing meta-reviews summarizing reviewers' discussions on ethical considerations.
    \item Consolidating research ethics committees from multiple conferences into a single board.
    \item Providing ethical training for reviewers.
    \item Introducing a “Profound Ethical Considerations” badge, similar to artifact evaluations.
    \item Making an ethics section mandatory for all research papers.
    \item Including external ethics experts in ethics boards at the conferences to ensure perspectives beyond our research community.
    \item Providing educational material on ethics in the form of categorized topics (e.g., human subjects,    responsible disclosure, etc.).
    \item Mentoring system for research ethics committee
\end{itemize}

\itembox{
Would you agree that the following instruments should be incorporated into the review forms for our conference? Please place the items shown below by drag \& drop into the response boxes depending on your assessment of the given item.

\emph{Ethics Score:} The ``score'' that the author receives in the review

\emph{Internal Ethics Concerns:} Internal documentation about the concerns for the REC and PC

\emph{Feedback for Author:} The feedback that the authors get with the review
}

\begin{itemize}
    \item Ethics Score: Binary (Ethical / Concerns)
    \item Ethics Score: Numeric
    \item Ethics Score: Severity (e.g., none, minor, moderate, severe)
    \item Internal Ethics Concerns / Considerations: Selectable Pre-formulated Categories
    \item Internal Ethics Concerns / Considerations: Free Text
    \item Feedback for Author: Pre-formulated Categories for Required Author's Action
    \item Feedback for Author: Free Text
    \item Reviewer Self-Assessment (Are you confident in evaluating the ethics of this paper)
    \item Ethics fields in review must be mandatory
\end{itemize}

\clearpage

\section{Process-Data Codebook}
\label{appx:process-data-codebook}

\revise{\Cref{tbl:process-data-codebook} shows the codes used to analyze ethics-related content in calls for papers and chair messages. The codes capture how conferences communicated ethical expectations and which ethics interventions they introduced.}

\begin{table}[h]
    \centering    
    \begin{tabularx}{\textwidth}{l X}
        \toprule
        Code & Comment \\[-1pt] \midrule
        Code of Conduct & 
        \revise{Reference to the conference's or organization's code of conduct.} \\[-1pt]
        Code of Ethics & 
        \revise{Reference to the conference's or organization's code of ethics.} \\[-1pt]
        Convince Reviewers & 
        \revise{It is mentioned that authors need to convince the reviewers that ethical standards were followed.}\\[-1pt]
        Ethical implications & 
        \revise{Ethical implications of research are mentioned in the CfP.} \\[-1pt]
        Ethics & \revise{This code is a structural code marking any mention of ethics-related information.}\\[-1pt]
        Ex-Post Ethics Board &
        \revise{It is mentioned that authors can reach out if they are unsure about the ethics of their research or seeking ways to reduce risk.} \\[-1pt]
        Human Subjects &
        \revise{It is mentioned that submissions with experiments on human subjects should discuss their ethical considerations.} \\[-1pt]
        In-paper Discussion &
        \revise{The sentence where it is explicitly stated that authors should discuss ethics in their paper.} \\[-1pt]
        IRB (or similar bodies) & 
        \revise{The conference asks authors to disclose whether they received an IRB approval or waiver.} \\[-1pt]
        Legal implications & 
        \revise{It is mentioned that experiments can have legal implications.} \\[-1pt]
        Mandatory ethics section & 
        \revise{It is mentioned that every submission must have an ethical considerations section.} \\[-1pt]
        Menlo Report &
        \revise{The Menlo report is mentioned as reference.} \\[-1pt]
        No-Change & as compared to the previous year (conference edition) \\[-1pt]
        Non-Public Data & 
        \revise{Analyzing non-public data derived from human subjects is mentioned as a research for which authors should consider ethics.} \\[-1pt]
        No-Go-Project & meaning that some projects should not be done for ethical reasons \\[-1pt]
        Research Ethics Committee &
        \revise{It is mentioned that the conferences have a REC.} \\[-1pt]
        Responsible Disclosure & \revise{It is mentioned that submissions need to discuss how they handled responsible disclosure}\\[-1pt]
        Right to reject &
        \revise{It is mentioned that the conference might reject submissions based on ethical concerns.} \\[-1pt]
        Risk-Benefit-Assessment &
        \revise{It is explicitly mentioned that the authors should assess and balance the risks and benefits of their research and minimize harm.} \\[-1pt]
        Sensitive Data & 
        \revise{It is mentioned that submissions need to discuss how they handled sensitive data.} \\[-1pt]
        \bottomrule
    \end{tabularx}
    \caption{Process-data Codebook}
    \label{tbl:process-data-codebook}
\end{table}

\section{Review Form Codebook}
\label{appx:review-codebook}

\revise{\Cref{tbl:review-codebook} shows the codes used to analyze ethics-related items in review forms. The codes capture whether and how reviewers were asked to identify, rate, or justify ethical concerns.}

\begin{table}[h]
    \centering  
    \begin{tabularx}{\textwidth}{l X}
        \toprule
        Code & Comment \\ \midrule
        Ethics & This code is a structural code marking any mention of ethics-related information. \\
        Ethics Binary & This code highlights instruments for a binary ethical rating, e.g., a flag. \\
        Ethics Rating & This code highlights instruments with a more fine-grained rating than binary, e.g., to indicate whether there are ethical concerns and how well they are addressed. \\
        Ethics Free Text & This code highlights free-text sections to justify ethical concerns. \\
        \bottomrule
    \end{tabularx}
    \caption{Review Form Codebook}
    \label{tbl:review-codebook}
\end{table}

\clearpage

\section{Interview Codebook}
\label{appx:interview-codebook}

\revise{\Cref{tbl:codebook} summarizes the codebook used for the interview analysis. It combines deductive codes derived from the interview guide with inductive codes added during coding.}

\begin{table}[h!]
    \begin{subtable}[t]{0.475\textwidth}
        \centering
        \subcaption{Deductive Codes}
        \begin{tabular}[t]{l}
            \toprule
            \textbf{Interview-Guide Code (IGC)} \\ \midrule
            IGC-01: Intro: Ethics General \\
            IGC-01.1: Ethics Feedback From Review \\ \midrule
            IGC-02: Ethics Review Procedure \\ 
            IGC-02.1: Existing Interventions \\
            \tabindent IGC-02.1.1: Ethics Frameworks \\
            \tabindent IGC-02.1.2: Ethics in Categories \\
            \tabindent IGC-02.1.3: IRB Disclosure \\
            \tabindent IGC-02.1.4: Mandatory Ethics Section \\
            \tabindent IGC-02.1.5: Research Ethics Committee \\
            \tabindent IGC-02.1.6: Responsible Disclosure Requirements \\
            \tabindent IGC-02.1.7: Right to Reject \\ \midrule
            IGC-03: Improvements \\ \midrule
            IGC-04: new-interventions\_own-ideas \\ \midrule
            IGC-05: New Interventions (survey tool 1) \\
            \tabindent IGC-05.1: Educational Material \\
            \tabindent IGC-05.2: Ethical Training / Education for PC \\
            \tabindent IGC-05.3: Ethics Badge \\
            \tabindent IGC-05.6: External Experts \\
            \tabindent IGC-05.7: Feedback to every paper \\
            \tabindent IGC-05.8: Mentoring for REC \\
            \tabindent IGC-05.9: Meta Reviews for Ethics Considerations \\ \midrule
            IGC-06: Review Form (survey tool 2) \\
            \tabindent IGC-06.1: Ethics fields mandatory \\
            \tabindent IGC-06.2: Ethics Score \\
            \tabindent IGC-06.2.1: Ethics-score\_binary \\
            \tabindent IGC-06.2.2: Ethics-score\_numeric \\
            \tabindent IGC-06.2.3: Ethics-score\_severity \\
            \tabindent IGC-06.4: Feedback to Authors on Ethics \\
            \tabindent IGC-06.4.1: Feedback to Authors on Ethics\_Free-Text \\
            \tabindent IGC-06.4.2: Feedback to Authors on Ethics\_Pre-Formulated \\
            \tabindent IGC-06.5: Internal Ethics Concerns \\
            \tabindent IGC-06.5.1: Internal Ethics Concerns\_Categories \\
            \tabindent IGC-06.5.2: Internal Ethics Concerns\_Free-Text \\
            \tabindent IGC-06.6: Self-Assessement \\ \midrule
            IGC-07: Reasoning \& History \\ \midrule
            IGC-08: Goals \\ 
            IGC: 08.1: Progress on ethics? \\ \midrule
            IGC-09: Importance: ethics vs. novelty \\ \midrule
            IGC-10: Aspects you don't like? \\ 
            \bottomrule
        \end{tabular}
        \label{codebook_deductive}
    \end{subtable}
    \begin{subtable}[t]{0.475\textwidth}
        \centering
        \subcaption{Inductive Codes}
        \begin{tabular}[t]{l}
            \toprule
            Goals of Ethics Practices \\
            \tabindent Consistency of ethical practies \\
            \tabindent Do Not Publish Unethical Papers \\
            \tabindent Education \\
            \tabindent Ethical Awareness \\
            \tabindent Ethics procedures = structure for reviewers \\
            \tabindent Goal is reached \\
            \tabindent Goal Not Reached Yet \\
            \tabindent Goals Unreachable \\
            \tabindent Improve papers \& make better science \\
            \tabindent Make ethics explicit \\
            \tabindent Making authors think aboutethics \\
            \tabindent Ongoing process \\
            \tabindent Positive impact on the world \\
            \tabindent pre-empting unethical studies \\
            \tabindent Preventing harm (ex-ante) \\
            \tabindent Protection from Liability \\
            \tabindent Public Reputation of Science \\
            \tabindent Signal to community \\
            \tabindent Systemize Ethics \\ \midrule
            Interventions \\
            \tabindent Artifacts Evaluations \\
            \tabindent Badge would be unfair \\
            \tabindent Checklist \\
            \tabindent Conference Guidelines \\
            \tabindent Contacting Law Enforecement \\
            \tabindent Ethical Training \\
            \tabindent Methods-specific Ethics \\
            \tabindent Ethics Frameworks \\
            \tabindent Ethics Section \\
            \tabindent Ethics Experts from other disciplines \\
            \tabindent IRBs are not sufficient \\
            \tabindent Mandatory Ethics Section \\
            \tabindent \tabindent Argument Against No-Ethics-Section-Needed \\
            \tabindent \tabindent Ethics Section is Not Hard to do \\
            \tabindent \tabindent Ethics Section Unclear \\
            \tabindent \tabindent Free ethics sections are hard to review \\
            \tabindent \tabindent Mandatory Ethics Sections Help Authors \\
            \tabindent \tabindent Not every paper needs to discuss ethics \\
            \tabindent \tabindent Vague Ethics Sections \\
            \tabindent Meta Reviews \\
            \tabindent Pre-Registration Panels \\
            \tabindent Rejection Because of Ethics \\
            \tabindent Research Ethics Committee \\
            \tabindent Responsible Disclosure \\
            \tabindent Review Form and Procedure \\
            \bottomrule
        \end{tabular}
        \label{codebook_inductive}
    \end{subtable}
    \caption{Interview Codebook}
    \label{tbl:codebook}    
\end{table}

\end{document}